\documentclass[aps,prl,twocolumn,showpacs,superscriptaddress,10pt,floatfix]{revtex4-2}

\usepackage{amsmath,amssymb,amsfonts}
\usepackage{graphicx}
\usepackage{bm}
\usepackage{physics}
\usepackage{siunitx}
\usepackage{hyperref}
\usepackage{color}
\usepackage{natbib}
\usepackage{dcolumn}
\newcommand{\sigmaAHE}{\sigma_{xy}}
\newcommand{\sigmaPHE}{\sigma_{xy}^{\mathrm{PHE}}}
\newcommand{\Bpar}{B_{\parallel}}
\newcommand{\Bperp}{B_{\perp}}
\newcommand{\RGg}{rhombohedral graphite}

\begin{document}

\preprint{APS/123-QED}

\title{Flat-Band Stoner Instability and Peierls-Phase Origin of the Transdimensional Anomalous Hall Effect in Rhombohedral Graphite}

\author{Yang Zhou}
\affiliation{Institute of Natural Sciences, Westlake Institute for Advanced Study, Hangzhou 310024, China}
\affiliation{Department of Physics, School of Science and Research Center for Industries of the Future, Westlake University, Hangzhou 310030, China}

\date{\today}

\begin{abstract}
A ``transdimensional'' anomalous Hall effect (TDAHE), in which both in-plane ($\Bpar$) and out-of-plane magnetic fields produce hysteretic Hall signals, was recently observed in nine-layer rhombohedral graphite~\cite{Li2026Nature}.
We present a microscopic theory attributing the TDAHE to a flat-band Stoner instability coupled to Peierls-phase gap modulation.
The large flat-band density of states satisfies a generalized Stoner criterion $U\rho(\varepsilon_F) > 1$~\cite{Bultinck2020}, driving a spin-valley-locked ferromagnet whose valley polarization $\eta$ breaks time-reversal symmetry and generates an intrinsic anomalous Hall conductivity (AHC).
The orbital $g$-factor $g_{\mathrm{orb}} = e d_0 v_F (N{-}1)/2 \propto (N{-}1)$, from the Peierls phase of interlayer hopping, then lets $\Bpar$ modulate the gap and produce the transdimensional response, which is proportional to the same $\eta$.
A self-consistent full $2N$-band Hartree-Fock calculation yields complete valley polarization ($\eta \to 1$) below a mean-field transition $T_c^{\mathrm{MF}} \approx 2.2$~K, reduced by 2D-Ising critical fluctuations to the experimental $T_c \approx 1.6$~K, with $R_{xy} \approx 1.5$~k$\Omega$.
Because both Hall responses are carried by one order parameter $\eta$, they share a single $T_c$, as observed; and $T_c$ is governed by the Stoner product $U\rho$ and is insensitive to the intervalley exchange, which only gates whether the valley-polarized phase forms---so neither the constrained-RPA nor the conventional screening value is a fitted parameter.
Beyond reproducing $R_{xy}$, $T_c$, and the phase window, the theory predicts a sharp onset of valley polarization between $N = 7$ and $N = 9$, a symmetry selection rule fixing the crescent Fermi surface to the $m{=}1$ nematic channel, and a transdimensional-to-conventional Hall ratio $\sigmaPHE/\sigmaAHE^{\mathrm{tot}} = g_{\mathrm{orb}}\Bpar/m$ that is independent of $\eta$ and of the interaction.
The $Z$-independence of the intrinsic AHC is verified within dynamical mean-field theory.
\end{abstract}


\maketitle

\section{Introduction}

The anomalous Hall effect (AHE)---the appearance of a transverse voltage in response to a longitudinal current without an external magnetic field---is a hallmark of time-reversal symmetry breaking in itinerant ferromagnets~\cite{Nagaosa2010}.
In two dimensions, the AHE acquires a topological character through the Berry curvature of Bloch bands, leading to the quantized anomalous Hall effect in magnetic topological insulators~\cite{Chang2013} and, more recently, in rhombohedral graphene moir\'e superlattices~\cite{Han2024Science,Chen2024}.
However, the conventional AHE is intrinsically a two-dimensional phenomenon: the Hall conductivity $\sigmaAHE$ couples only to the \textit{out-of-plane} component of orbital magnetization through the Berry curvature in the plane of electron motion.

Li \textit{et al.}~\cite{Li2026Nature} overturned this picture in a nine-layer \RGg{} device.
They observed a ``transdimensional'' anomalous Hall effect (TDAHE) in which \textit{both} out-of-plane ($\Bperp$) and \textit{in-plane} ($\Bpar$) magnetic fields produce hysteretic Hall resistance.
The in-plane coercive field ($160$--$500$~mT) is two orders of magnitude larger than the out-of-plane one ($\sim 3$~mT), yet both signals share the same temperature and gate-voltage dependence, pointing to a common origin.
Both $\Bperp$ and $\Bpar$ hysteresis vanish at the same $T_c \sim 1.6$~K, establishing that the TDAHE exists only within the ferromagnetic phase~\cite{Li2026Nature}.

The theoretical understanding of the TDAHE remains incomplete. Li \textit{et al.}~\cite{Li2026Nature} reported Hartree-Fock calculations showing a crescent-shaped Fermi surface with in-plane orbital magnetization $\sim 2.2\,\mu_B$/electron but deferred a quantitative theory. Zheng \textit{et al.}~\cite{Zheng2025NanoLett} proved that interlayer coherence can produce an intrinsic planar Hall effect through the Berry-connection susceptibility, but their non-interacting framework imposes symmetry breaking by hand---it has no order parameter, no spontaneous transition, and therefore no $T_c$, layer-number threshold, or gate-tuned onset, all of which are observed~\cite{Li2026Nature}. DMFT work~\cite{Georges1996} established that the intrinsic AHC is independent of the quasiparticle weight $Z$~\cite{Acheche2018}, ruling out any $1/Z^2$ enhancement.

Here we develop a microscopic theory of the TDAHE based on a flat-band Stoner instability and orbital Zeeman coupling. In rhombohedral graphite the large DOS drives a generalized Stoner instability~\cite{Bultinck2020} that simultaneously polarizes spin and valley; the resulting ferromagnet breaks TRS and generates Berry curvature, while the orbital Zeeman coupling $g_{\mathrm{orb}} \propto (N{-}1)$, arising from the Peierls phase of interlayer hopping, lets in-plane fields modulate the gap. Our treatment supplies what Ref.~\onlinecite{Li2026Nature} defers: (i) a single-order-parameter framework tying both Hall responses to one valley polarization $\eta$, so their common $T_c$ follows as a structural consequence; (ii) the finding that $T_c$ is governed by the Stoner product $U\rho$ and is insensitive to the intervalley exchange, which only gates whether the valley-polarized phase forms---so the exchange strength is not a fitted parameter; (iii) a symmetry selection rule fixing the crescent to the $m{=}1$ nematic channel; and (iv) a prediction for the transdimensional-to-conventional Hall ratio $\sigmaPHE/\sigmaAHE^{\mathrm{tot}} = g_{\mathrm{orb}}\Bpar/m$, set by the single-particle gap and Peierls $g$-factor and so independent of $\eta$ and the interaction.

\section{Model}

\textit{Tight-binding Hamiltonian.}---
We consider $N$-layer ABC-stacked rhombohedral graphene with the Slater-Koster parameterization of Ref.~\onlinecite{Mikheeva2025}.
The full Hamiltonian in the sublattice basis $\{A_1, B_1, \ldots, A_N, B_N\}$ is:
\begin{equation}
  H(\mathbf{k}) = H_{\mathrm{intra}}(\mathbf{k}) + H_{\mathrm{inter}} + H_D,
  \label{eq:hamiltonian}
\end{equation}
where $H_{\mathrm{intra}}$ describes intralayer graphene with hopping $\gamma_0 = 3.16$~eV, $H_{\mathrm{inter}}$ includes interlayer hoppings $\gamma_1 = 0.39$~eV (dominant), $\gamma_3 = 0.315$~eV (trigonal warping), and $H_D$ is the displacement field.
Near the $K$ point, the low-energy physics is captured by a two-band effective Hamiltonian~\cite{Mikheeva2025}.
Defining $\mathbf{p} = \hbar(\mathbf{k} - \mathbf{K})$ as the momentum measured from $K$ and $P = \xi p_x - ip_y$ with valley index $\xi = \pm 1$:
\begin{equation}
  H_{\mathrm{eff}} = m\sigma_z + v_{\mathrm{eff}} P^N \sigma_+ + v_{\mathrm{eff}}^{*} P^{*N} \sigma_-,
  \label{eq:2band}
\end{equation}
where $m = eDd_0/2$ is the displacement-field-induced gap and $v_{\mathrm{eff}} = v_F^N/\gamma_1^{N-1}$ is the effective velocity. Trigonal warping ($\gamma_3$) enters at $O(p^{N-3})$ and is omitted from the projected two-band form (all analytic results are to leading order in $p^N$); the full $\gamma_3$-containing tight-binding Hamiltonian is retained for all numerics (Supplementary).

\textit{Orbital $g$-factor.}---
The flat bands carry a large orbital magnetic moment from the Peierls phase accumulated across the rhombohedral stack. The out-of-plane moment at momentum $\mathbf{p}$, from the standard Berry-curvature orbital moment~\cite{Xiao2010} on Eq.~\eqref{eq:2band}, is:
\begin{equation}
  \mathcal{M}_z(\mathbf{p}) = \frac{e v_F d_0 (N-1)}{2} \cdot \frac{m}{\sqrt{m^2 + v_{\mathrm{eff}}^2 p^{2N}}},
  \label{eq:orb_mag}
\end{equation}
where $d_0 = 0.335$~nm is the interlayer spacing.
The on-site ($\mathbf{p} = 0$) value defines an orbital $g$-factor [Eq.~\eqref{eq:g_orb}]:
\begin{equation}
  g_{\mathrm{orb}} = \frac{e \, d_0 \, v_F \, (N-1)}{2},
  \label{eq:g_orb}
\end{equation}
giving the gap shift $\delta m = g_{\mathrm{orb}} \Bpar$ from the Peierls-phase modification of interlayer hopping. For $N = 9$ this yields $g_{\mathrm{orb}} = 1.37$~meV/T ($g_{\mathrm{orb}}/\mu_B \approx 24$), an order of magnitude larger than typical spin $g$-factors; the $g_{\mathrm{orb}} \propto (N-1)$ scaling reflects the extended orbital structure across the stack.

\textit{Stoner instability.}---
The flat bands near the Fermi level produce a strongly enhanced DOS.
For the two-band model, $\rho(\varepsilon_F) \propto 1/(v_{\mathrm{eff}} \, p_F^{N-1})$ is strongly enhanced near the flat-band edge; the full $2N$-band calculation gives $\rho(\varepsilon_F) \approx 3.56$~states/eV per unit cell for $N = 9$ (the two-band van~Hove singularity overestimates this as $\sim\!10$), two orders of magnitude larger than monolayer graphene.
The Hubbard $U$ drives a Stoner instability~\cite{Bultinck2020} when:
\begin{equation}
  U \, \rho(\varepsilon_F) > 1,
  \label{eq:stoner}
\end{equation}
the generalized Stoner criterion for multivalley ferromagnetism.
Once satisfied, the system develops exchange splitting $\Delta_{\mathrm{ex}} = UM(T)$ that onsets at $T_c$, breaking time-reversal symmetry and generating Berry curvature.
The intervalley exchange that drives the valley polarization is $V_{\mathrm{iv}}/U \approx 1$ at the constrained-RPA value, far above the graphene intravalley $\sim\!0.2$--$0.3$~\cite{Schuler2013,Wehling2011} because it scatters at $|\mathbf{K}{-}\mathbf{K}'| \gg 2k_F$ where the flat-band channel is nearly unscreened ($\varepsilon_{\mathbf{K}-\mathbf{K}'}\!\approx\!1$); its microscopic origin is detailed in the Results and Supplementary.

\section{Berry curvature and transport}

\textit{Intrinsic AHE from ferromagnetic order.}---
The Berry curvature of the two-band model (lower band) is~\cite{Slizovskiy2019,Mikheeva2025}:
\begin{equation}
  \Omega(\mathbf{p}) = \frac{N^2 m v_{\mathrm{eff}}^2 p^{2N-2}}{2\left(m^2 + v_{\mathrm{eff}}^2 p^{2N}\right)^{3/2}},
  \label{eq:berry}
\end{equation}
where the $N^2$ factor originates from the velocity operator $v_{x,y} = \partial H/\partial p_{x,y} \propto N P^{N-1}$ appearing twice in the Kubo formula.
In the ferromagnetic state, the gap $m$ in Eq.~\eqref{eq:berry} is the total gap containing both the displacement-field contribution and the exchange splitting: $m = m_D + \Delta_{\mathrm{ex}}$.
The intrinsic anomalous Hall conductivity (per valley per spin) is:
\begin{equation}
  \sigmaAHE^{\mathrm{int}} = \frac{e^2}{h} \cdot \frac{Nm}{2\varepsilon_F},
  \label{eq:sigma_int}
\end{equation}
exhibiting the characteristic $N$-fold enhancement~\cite{Mikheeva2025}.
For $N = 9$ in the flat-band limit ($\varepsilon_F \approx m$), this gives $\sigmaAHE^{\mathrm{int}} = N/2 = 4.5\,e^2/h$ (per valley per spin).

\textit{Valley polarization and net AHC.}---
The Berry curvature at $K$ and $K'$ has opposite sign ($\Omega_{K} = -\Omega_{K'}$), so spin-dependent exchange splitting alone does not lift the cancellation: $\sigmaAHE^{K,\uparrow} + \sigmaAHE^{K',\uparrow} = 0$ within each spin channel. A net AHC therefore requires \textit{valley polarization}---an imbalance $n_K \neq n_{K'}$ at the Fermi level---which the flat-band Stoner instability drives simultaneously with spin through intervalley exchange~\cite{Bultinck2020}, producing a spin-valley-locked ferromagnetic state (consistent with the $\sim 2.2\,\mu_B$/electron orbital magnetization of Ref.~\onlinecite{Li2026Nature}). Parametrizing the valley imbalance by $\eta \in [0,1]$, the total AHC is:
\begin{equation}
  \sigmaAHE^{\mathrm{tot}} = \eta \cdot g_s \cdot \sigmaAHE^{\mathrm{int}},
  \label{eq:sigma_tot}
\end{equation}
where $g_s = 2$ counts the spin channels (both polarized). Since $\eta(T)$ is part of the ferromagnetic order parameter and $\eta(T) \to 0$ as $T \to T_c$, the total AHC vanishes continuously at the transition.

\textit{Peierls-phase (orbital Zeeman) mechanism for the TDAHE.}---
Having established the valley-polarized AHC, we now show how in-plane fields produce a transdimensional response.
The in-plane field couples to the system through the Peierls phase of interlayer hopping, which is layer-resolved across the rhombohedral stack (Supplementary).
Although the out-of-plane moment $\mathcal{M}_z$ [Eq.~\eqref{eq:orb_mag}] couples only to $B_\perp$ (so $\mathcal{M}_z \cdot B_\parallel = 0$), the same Peierls phase generates an in-plane-field response through a different channel: $B_\parallel$ accumulates a layer-dependent hopping-phase shift whose two-band projection shifts the gap.
For $\Bpar$ along $x$, this shifts the gap to leading order:
\begin{equation}
  m \to m + g_{\mathrm{orb}} \Bpar,
  \label{eq:gap_shift}
\end{equation}
This shift [Eq.~\eqref{eq:gap_shift}] applies identically to both valleys (the orbital Zeeman coupling is independent of $\xi$), so the valley-polarized AHC acquires a modulation:
\begin{equation}
  \sigmaPHE = \frac{\partial \sigmaAHE^{\mathrm{tot}}}{\partial m} \cdot g_{\mathrm{orb}} \Bpar
  = \eta \, g_s \, \frac{e^2}{h} \cdot \frac{N \, g_{\mathrm{orb}} \Bpar}{2\varepsilon_F},
  \label{eq:sigma_phe}
\end{equation}
where we used $\sigmaAHE^{\mathrm{int}} \propto m$ at fixed $\varepsilon_F$.
The \textit{enabling factor} $\eta(T)$ appears in both $\sigmaAHE^{\mathrm{tot}}$ and $\sigmaPHE$: the transdimensional response is proportional to the valley polarization, which tracks the ferromagnetic order parameter:
\begin{equation}
  \sigmaPHE(T) = \eta(T) \cdot g_s \cdot \frac{e^2}{h} \cdot \frac{N \, g_{\mathrm{orb}} \Bpar}{2\varepsilon_F}.
  \label{eq:sigma_phe_M}
\end{equation}
Above $T_c$, $\eta = 0$ and both $\sigmaAHE^{\mathrm{tot}}$ and $\sigmaPHE$ vanish, consistent with the experimental observation that the TDAHE disappears at $T_c$.
At $\Bpar = 0.3$~T and $m = 142$~meV, the ratio of the transdimensional to conventional AHC is $\sigmaPHE/\sigmaAHE^{\mathrm{tot}} = g_{\mathrm{orb}}\Bpar/m \approx 0.3\%$, independent of $\eta$.
This small ratio reflects the weakness of the orbital Zeeman coupling relative to the gap, not the ferromagnetic order itself.

\section{Results}

\begin{figure}[!ht]
\includegraphics[width=\columnwidth]{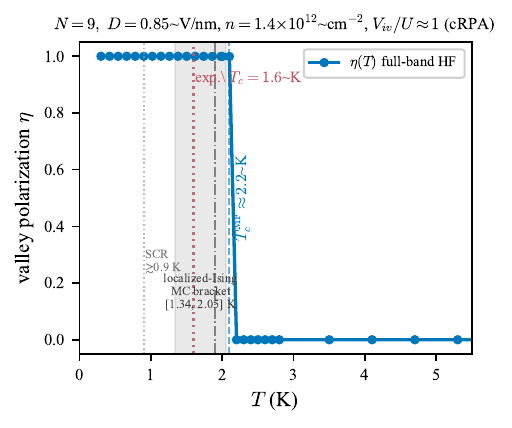}
\caption{Self-consistent valley polarization $\eta(T)$ from the full $2N$-band Hartree-Fock calculation ($N=9$, $D=0.85$~V/nm). The valley order saturates at $\eta\to 1$ and vanishes at $T_c^{\mathrm{MF}}\approx 2.2$~K (blue dashed). Shaded band: localized-Ising $T_c$ bracket $[1.34, 2.05]$~K (NN lower bound, bare-$1/r$ upper bound; intermediate ranges give $\approx\!1.9$~K, dash-dot, Supplementary). Red dotted: experimental $T_c=1.6$~K~\cite{Li2026Nature}; grey dotted: SCR lower bound $\gtrsim\!0.9$~K.}
\label{fig:stoner}
\end{figure}

\textit{Self-consistent valley polarization.}---
The flat-band Stoner instability ($U\rho > 1$) drives a spin-polarized ferromagnetic (FM) ground state for $N \ge 5$. At $N = 9$ and $D = 0.85$~V/nm the intervalley exchange ($V_{\mathrm{iv}}/U \approx 1$) selects the spin-valley-locked ferromagnet (SVM) with valley polarization $\eta > 0$ as the self-consistent ground state (Fig.~\ref{fig:stoner}); the spin-only FM remains a local minimum, with the SVM free-energy-lower whenever $V_{\mathrm{iv}}/U$ exceeds the threshold $\sim\!0.65$ (Discussion).
The full $2N$-band Hartree-Fock calculation at the cRPA value $V_{\mathrm{iv}}/U \approx 1$ gives valley polarization saturating to $\eta \to 1$ in the ordered phase and a mean-field transition $T_c^{\mathrm{MF}} \approx 2.2$~K (set by $U\rho \approx 2.2$ per spin-valley channel, Fig.~\ref{fig:stoner}); the experimental $T_c \approx 1.6$~K~\cite{Li2026Nature} is lower by a factor $\sim\!1.4$, attributed to strong critical fluctuations of the flat band (placing the $\mathbb{Z}_2$ valley transition in the 2D Ising regime, Discussion). This saturated prediction $\eta_{\rm HF} \to 1$ admits an independent, model-independent cross-check: the orbital magnetization $\sim 2.2\,\mu_B$/electron reported by Li~\textit{et al.}~\cite{Li2026Nature} for $N = 9$ (with $T_c \sim 1.6$~K) and the per-particle band-edge moment $g_{\mathrm{orb}}/\mu_B \approx 24$ give a ratio $2.2/24 \approx 0.1$ that sets a weak \emph{lower bound} on $\eta$; the inequality $\eta_{\rm HF} = 1 \geq 0.1 = \eta_{\rm bound}$ is a consistency check, not a contradiction.
The SVM ground state is found for $N \ge 8$ ($\eta \to 1$, Table~\ref{tab:scaling}); for $N = 5$ and $7$, though still ferromagnetic ($U\rho \gg 1$), the valley susceptibility stays below threshold and the ground state remains spin-only FM ($\eta = 0$). Table~\ref{tab:scaling} summarizes the predicted quantities across layer numbers.
\begin{table}[ht]
\caption{Predicted quantities for $N$-layer rhombohedral graphite at $D = 0.85$~V/nm ($m = 142$~meV).
Here $g_{\mathrm{orb}} = e d_0 v_F (N{-}1)/2$ and $\sigma^{\mathrm{int}} = N/2$ (per valley per spin, flat-band limit).
The Stoner product $U\rho(\varepsilon_F)$ (summed over the four spin/valley flavors, i.e.\ per unit cell; $\approx 2.2$ per spin-valley channel) is from the full $2N$-band tight-binding Hamiltonian with thermal broadening $k_BT$.
Valley polarization $\eta$ and $R_{xy} = \sigma^{\mathrm{tot}}/(\sigma_{xx}^2 + \sigma^{\mathrm{tot}\,2}) \times h/e^2$ from the self-consistent full $2N$-band Hartree-Fock calculation at the constrained-RPA value $V_{\mathrm{iv}}/U \approx 1$, deep in the ordered phase ($T = 0.5$~K, where $\eta$ is saturated), with $\sigma_{xx} = 8\,e^2/h$.}
\label{tab:scaling}
\begin{ruledtabular}
\begin{tabular}{cccccc}
$N$ & $g_{\mathrm{orb}}$ (meV/T) & $U\rho(\varepsilon_F)$ & $\eta$ & $\sigma^{\mathrm{tot}}$ ($e^2/h$) & $R_{xy}$ (k$\Omega$) \\
\hline
3 & 0.34 & 0.55 & 0 & 0 & 0 \\
5 & 0.69 & 7.31 & 0 & 0 & 0 \\
7 & 1.03 & 8.90 & 0 & 0 & 0 \\
8 & 1.20 & 8.88 & 1.00 & 8.0 & 1.6 \\
9 & 1.37 & 8.89 & 1.00 & 9.0 & 1.6 \\
11 & 1.71 & 8.88 & 1.00 & 11.0 & 1.5 \\
\end{tabular}
\end{ruledtabular}
\end{table}

\begin{figure}[t]
\includegraphics[width=\columnwidth]{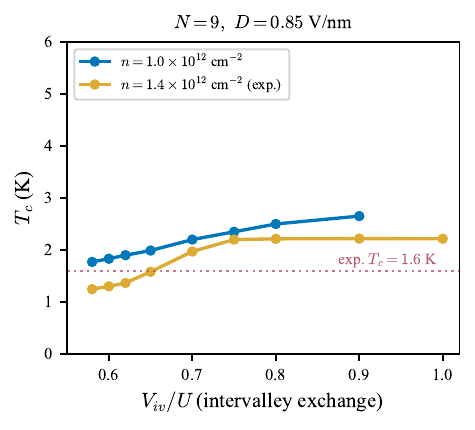}
\caption{Insensitivity of $T_c$ to the intervalley exchange. Self-consistent $2N$-band Hartree-Fock $T_c^{\mathrm{MF}}(V_{\mathrm{iv}}/U)$ for $N=9$ at $D=0.85$~V/nm, at the experimental density $n=1.4\times10^{12}$~cm$^{-2}$ (orange) and a lower density (blue). $T_c$ saturates at $\approx\!2.2$~K for $V_{\mathrm{iv}}/U \gtrsim 0.75$; below $\sim\!0.65$ the valley polarization is lost. Red dotted: experimental $T_c=1.6$~K~\cite{Li2026Nature}.}
\label{fig:tcviv}
\end{figure}

\textit{Common $T_c$ for AHE and TDAHE.}---
\textit{Layer-number onset and intervalley exchange.}---
The onset of valley polarization for $N \ge 8$ (but not $N \le 7$) holds across the cRPA/conventional range: at $V_{\mathrm{iv}}/U \approx 1$ the $N = 9$ device sits deep in the SVM phase ($\eta \to 1$, Table~\ref{tab:scaling}), with $N = 8$ at the onset boundary, and the predicted $R_{xy} \approx 1.5$--$1.6\,\mathrm{k}\Omega$ is consistent with the experimental $R_{xy} \approx 1.4\,\mathrm{k}\Omega$. The microscopic origin of the exchange is the projected Coulomb interaction in the flat-band subspace: the K and K$'$ flat bands are both localized on the A$_1$ site, so the bare exchange $\langle\psi_K,\psi_{K'}|V|\psi_{K'},\psi_K\rangle \approx U$ (a full-band Fock projection gives $2.53$~eV $\approx U = 2.5$~eV), unlike ideal graphene where sublattice complementarity forces it to zero~\cite{Schuler2013,Wehling2011}. Because the intervalley channel scatters at $|\mathbf{K}{-}\mathbf{K}'| \gg 2k_F$ where the flat-band intraband polarizability vanishes, only the 16 remote bands screen it, and a like-for-like constrained-RPA comparison gives $V_{\mathrm{iv}}/U \approx 1$ (Supplementary); the conventional momentum-independent screening instead reproduces the graphene value $\sim\!0.3$. Crucially, $V_{\mathrm{iv}}/U$ controls only whether the SVM forms (threshold $\sim\!0.65$), not $T_c$, which is set by $U\rho \gg 1$: scanning $V_{\mathrm{iv}}/U$ (Fig.~\ref{fig:tcviv}), $T_c^{\mathrm{MF}}$ saturates at $\approx\!2.2$~K for $V_{\mathrm{iv}}/U \gtrsim 0.75$, so both the cRPA ($\approx\!1$) and conventional ($\approx\!0.3$) values place $N=9$ on the same ordered side with the same $T_c$.

\textit{Density dependence.}---
\begin{figure}[ht]
\includegraphics[width=\columnwidth]{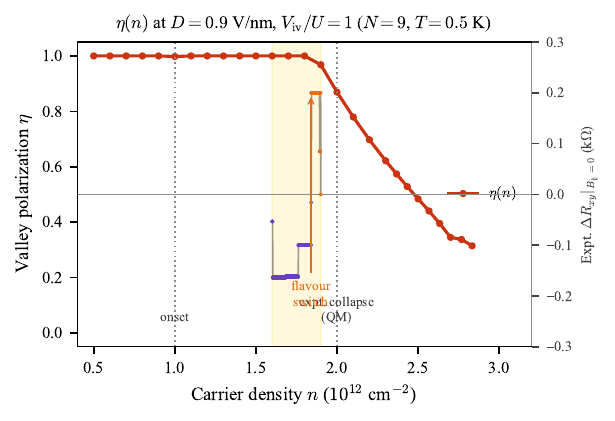}
\caption{Self-consistent valley polarization $\eta(n)$ (left axis, red) at $D = 0.9$~V/nm ($N = 9$, $T = 0.5$~K), saturated ($\eta\to 1$) for $n \lesssim 1.8\times10^{12}$~cm$^{-2}$. Right axis: experimental $\Delta R_{xy}$ at $B_{\parallel}=0$ (digitized from Extended Data Fig.~6 of Ref.~\onlinecite{Li2026Nature}), purple/orange for $\Delta R_{xy}\lessgtr 0$. $\eta$ is the order-parameter \emph{magnitude}, so it stays saturated across the switching window ($n=1.6$--$1.9$) while the $\Delta R_{xy}$ \emph{sign} reverses at $n\approx 1.83$, marking the valley flavour switch $K\leftrightarrow K'$ at fixed $|\eta|$. Yellow band: switching window; dotted lines mark hysteresis onset ($n=1.0$) and loop collapse ($n=2.0$, into the quarter-metal phase).}
\label{fig:phase}
\end{figure}
Figure~\ref{fig:phase} shows $\eta(n)$ at $D = 0.9$~V/nm, the experimental optimal point used for the density sweep (the $T_c$ and Table~\ref{tab:scaling} values are computed at the nearby $D = 0.85$~V/nm; the band-edge DOS varies by only $\sim\!2\%$ between them): the spin-polarized FM is the self-consistent baseline everywhere, and valley polarization ($\eta \to 1$, SVM) develops as an additional ordering for $n \lesssim 1.8 \times 10^{12}$~cm$^{-2}$, where the Fermi level sits at the flat-band edge and the DOS is maximal. The valley order is controlled primarily by density, not displacement field (the band-edge DOS varies by only $\sim\!2\%$ across $D = 0.7$--$1.0$~V/nm), consistent with the broad experimental $D$ window. Within the experimental density window ($n = 1.0$--$2.0 \times 10^{12}$~cm$^{-2}$) the reported ``electrical switching of the magnetic order''~\cite{Li2026Nature}---a sign reversal of $\Delta R_{xy}$ at $n \approx 1.83$---is a first-order switch between the two degenerate SVM solutions ($\pm\eta$), i.e.\ the ordered valley flavour changing $K \leftrightarrow K'$; our continuous mean-field $\eta$ tracks only the magnitude $|\eta|$, which stays saturated through the switch, so the saturated plateau and the observed sign reversal are consistent rather than contradictory. The experimental hysteretic loop collapses sharply at $n \approx 2.0$ into the quarter-metal phase, earlier and sharper than the continuous $\eta(n) \to 0$ near $n \sim 3.5$, consistent with a first-order Fermi-surface-topology transition at the quarter-metal--TDAHE boundary that a continuous order parameter cannot capture.

\textit{Correlation independence.}---
One might expect DMFT quasiparticle renormalization ($Z<1$) to enhance the intrinsic AHC through a $1/Z^2$ factor. This is ruled out: the $Z$ in the Kubo formula cancels exactly against band renormalization~\cite{Acheche2018}; a Hubbard-I DMFT calculation for $N=9$ confirms $\sigmaAHE^{\mathrm{int}}(Z)/\sigmaAHE^{\mathrm{int}}(\mathrm{bare}) = 1$ within $10^{-8}$ for $U=0.5$--$5.0$~eV (Supplementary), so the TDAHE is governed by the Stoner ferromagnetism ($\eta$) and the orbital Zeeman coupling ($g_{\mathrm{orb}}$), not by correlation enhancement.

\section{Discussion}
\label{sec:discussion}

The TDAHE requires two simultaneous conditions: (i) valley polarization $\eta > 0$, which breaks TRS and generates a net AHC; and (ii) the orbital Zeeman coupling $g_{\mathrm{orb}}$, which lets in-plane fields modulate the gap.
The bare planar Hall conductivity ($\sim 0.3\%$ of $\sigmaAHE^{\mathrm{int}}$ at $B = 0.3$~T) is small, but ferromagnetic order renders it observable by breaking TRS.
Because both $\sigmaAHE \propto \eta$ and $\sigmaPHE \propto \eta$ are carried by the same valley order parameter, they necessarily vanish together at the single temperature where $\eta(T) \to 0$; the common $T_c$ is thus a structural consequence of the single-order-parameter framework rather than an additional assumption, and is consistent with the experimental observation that both Hall signals disappear together~\cite{Li2026Nature}.

\textit{Mean-field $T_c$ and critical fluctuations.}---
The cRPA Hartree-Fock calculation gives $T_c^{\mathrm{MF}} \approx 2.2$~K for $N=9$, exceeding the experimental $T_c \approx 1.6$~K by a factor $\sim\!1.4$---precisely what critical-fluctuation theory predicts for a flat-band system. The valley polarization is a $\mathbb{Z}_2$ (Ising) order parameter, so the transition evades the Mermin-Wagner theorem and a finite $T_c$ is permitted. The flat-band stiffness $c_v$ is vanishingly small (the band disperses by only $\sim\!0.01$~meV, and the huge flat-band DOS drives $q_{\mathrm{TF}} \sim 350$~nm$^{-1}$, screening the Coulomb to a contact), giving a coherence length $\xi_0 \approx 0.35$~nm $\ll N d_0 = 3.0$~nm: the transition is two-dimensional and governed by the 2D Ising universality class rather than mean-field theory. We bracket the corrected $T_c$: a classical Ising Monte Carlo of the projected $\mathbb{Z}_2$ valley model---exchange range bounded by bare-$1/r$ from above and $\xi_0$ from below---gives $T_c \in [1.34, 2.05]$~K ($\approx\!1.9$~K for intermediate ranges), with SCR theory providing a Gaussian lower bound $\gtrsim 0.9$~K; the experimental $1.6$~K falls inside this bracket, in its lower portion (Supplementary). The factor-$\sim\!1.4$ suppression of $T_c^{\mathrm{MF}}$ noted above and the lower-portion placement here are two views of the same physics: critical fluctuations suppress $T_c$ below the mean-field value, and the effective valley-exchange range---shorter than the bare $1/r$ because the direct Coulomb does not couple valleys---places the localized-Ising $T_c$ in the lower part of the bracket. A sign-free determinant QMC of the two-valley Hubbard model confirms the correlated flat-band regime and 2D Ising universality but, on the minimal uniform-repulsion lattice, develops no long-range valley order at the mean-field scale; the momentum-dependent intervalley exchange needed for an unbiased $T_c$ is now sign-free-tractable and left to future work.

\textit{Crescent Fermi surface: a mean-field feature decoupled from the Hall response.}---
Li \textit{et al.}~\cite{Li2026Nature} reported an unrestricted Hartree-Fock (HF) crescent-shaped, $C_3$-broken Fermi surface (FS), but this geometry and its in-plane orbital magnetization are mean-field predictions, not directly measured quantities. Our theory instead targets the uniform valley polarization $\eta$. Allowing a $C_3$-breaking nematic self-energy on the flat-band surface orbital,
\begin{equation}
  \Sigma_{A_1}(\mathbf{k}) = \Delta\,\cos\theta_{\mathbf{k}},
  \label{eq:crescent}
\end{equation}
in the $m{=}1$ angular channel of the intervalley Coulomb interaction, we recover the crescent as a second-order instability at $V_1 > V_{1c} \approx 1.7$~eV (analytic; $\approx\!2.8$~eV numerically), with $\Delta \propto (V_1 - V_{1c})^{1/2}$. Symmetry selects the $m{=}1$ channel: a diagonal exchange is linear in the order parameter and distorts the ring into a crescent, whereas off-diagonal intervalley coherence enters at quadratic order and can only yield the $m{=}2$ ellipse---a selection rule absent from Ref.~\onlinecite{Li2026Nature}. Crucially, the full-band Kubo $\sigmaAHE$ shifts by only $1.6\%$ between the crescent and the ring ($0.495$ vs $0.487~e^2/h$ per valley/spin, Supplementary), so the consistency of our $\sigmaAHE$ with the measured $R_{xy}$ is insensitive to the FS geometry; a quantitative in-plane orbital magnetization and the in-plane coercive fields ($160$--$500$~mT) lie beyond equilibrium mean field.

Two features distinguish the Stoner mechanism from purely orbital or interlayer-coherence frameworks~\cite{Zheng2025NanoLett}: the TDAHE appears only when $U\rho(\varepsilon_F) > 1$, predicting a sharp onset at the correlated-regime boundary, as observed~\cite{Zhang2024PNAS,Li2026Nature}; and the Peierls-phase $g_{\mathrm{orb}} \propto (N-1)$ is verified by full tight-binding diagonalization. The SVM is the global minimum below $T_c$, with the spin-only FM a metastable local minimum, so sweeping $D$ or $n$ across the threshold can switch between the two minima, predicting hysteresis consistent with the electric-field hysteresis near the quarter-metal--TDAHE boundary~\cite{Li2026Nature}.

The Chern-number reversal under in-plane fields in eight-layer rhombohedral graphite~\cite{Zan2026} involves related physics but falls outside our two-band framework: our model predicts $B_{\mathrm{rev}} \sim m/g_{\mathrm{orb}} \approx 118$~T for $N = 8$, versus $\sim\!0.4$~T experimentally, indicating that the moir\'e superlattice there qualitatively modifies the physics (spin-orbit-assisted band inversion or interaction-driven level crossing), left to future work.

\textit{Testable predictions.}---

(1) {\bf Layer-number scaling} (Table~\ref{tab:scaling}): valley polarization onsets sharply at $N = 8$, with $\eta$ jumping from $0$ ($N{=}7$) to $1$ ($N{=}8$), leaving $N = 8$ as a testable boundary layer with the experimental $N = 9$ device well inside the ordered phase; the intrinsic AHC $\sigma^{\mathrm{tot}} \propto N$ ($\approx\!11\,e^2/h$ for $N{=}11$) and $g_{\mathrm{orb}} \propto (N{-}1)$ then grow with $N$, while $N = 3$ ($U\rho = 0.55 < 1$) stays outside the correlated regime.

(2) {\bf Doping-switched TDAHE}: At $D = 0.9$~V/nm, $N = 9$, the valley order weakens continuously beyond $n \approx 1.8 \times 10^{12}$~cm$^{-2}$, dropping below $\eta = 0.5$ near $n \approx 2.5 \times 10^{12}$~cm$^{-2}$ (Fig.~\ref{fig:phase}). The experimental hysteretic loop collapses sharply at $n \approx 2.0 \times 10^{12}$~cm$^{-2}$ into the quarter-metal phase; the predicted continuous $\eta(n)$ decrease and the experimentally observed switching of the magnetic order near this boundary (Extended Data Fig.~6 of Ref.~\onlinecite{Li2026Nature}) can be verified by sweeping the back gate at fixed $\Bpar = 0.3$~T.

(3) {\bf Common $T_c$ and interaction-independent amplitude ratio}: Both $R_{xy}(\Bperp)$ and $R_{xy}(\Bpar)$ hysteresis vanish at the same $T_c$, because $\eta(T) \to 0$ controls both signals. Their ratio is set by the single-particle band structure rather than by the ordered state, $\sigmaPHE/\sigmaAHE^{\mathrm{tot}} = g_{\mathrm{orb}}\Bpar/m \approx 0.3\%$ at $\Bpar = 0.3$~T, independent of $\eta$, of $U$, and of temperature---so it can be tested without knowing the valley polarization or the interaction strength.

(4) {\bf Magnetization switching}: near the quarter-metal--TDAHE boundary the observed sign reversal of $\sigma_{xy}$~\cite{Li2026Nature} corresponds in our framework to the valley polarization flipping $\eta \to -\eta$ (majority valley $K \leftrightarrow K'$); the two degenerate SVM solutions have identical free energies, and a field or gate sweep selects one, producing the hysteresis loop with sign reversal.

(5) {\bf In-plane isotropy}: The TDAHE amplitude and coercive field are isotropic under rotation of $\Bpar$ in the sample plane, because $\mathcal{M}_z$ is always perpendicular to the layers.

(6) {\bf Moir\'e enhancement}: In moir\'e rhombohedral graphite~\cite{Liu2025twist}, the mini-band DOS enhancement ($\sim a_{\rm moire}^2/a^2 \sim 10^2$) should raise $T_c$ and expand the TDAHE window.

\section{Conclusion}

We have presented a microscopic theory of the TDAHE rooted in a flat-band Stoner instability ($U\rho > 1$) and Peierls-phase orbital Zeeman coupling. The Stoner order parameter $\eta$ breaks TRS and generates Berry curvature ($\sigmaAHE^{\mathrm{int}} = (e^2/h)Nm/2\varepsilon_F$), while $g_{\mathrm{orb}} \propto (N{-}1)$ lets in-plane fields modulate the gap ($\sigmaPHE = \eta g_s (e^2/h) N g_{\mathrm{orb}}\Bpar/2\varepsilon_F$). Because $\eta(T)$ controls both Hall responses, they share a single $T_c$, as observed. The essential advance over the non-interacting planar Hall framework~\cite{Zheng2025NanoLett} is the emergence of the order itself: $\eta$ is a self-consistent interaction-driven order parameter with a Stoner threshold, a spontaneous onset, and a critical temperature, none of which exist when symmetry breaking is imposed by hand---which is why the two responses lock to one $T_c$, the TDAHE appears only above a layer-number threshold, and the amplitude ratio $\sigmaPHE/\sigmaAHE^{\mathrm{tot}} = g_{\mathrm{orb}}\Bpar/m$ is set entirely by the single-particle gap and $g$-factor, independent of $\eta$ and the interaction.

\begin{acknowledgments}
This work was supported by the Scientific Research Project (No.WU2025B011) and the Start-up Funding of Westlake University.
\end{acknowledgments}

\section*{Data and Code Availability}
All numerical data and source code supporting this study are available from the corresponding author upon reasonable request and will be deposited in a public repository upon publication. During the preparation of this work the authors used Claude (Anthropic) to improve the language and readability of the manuscript; the authors reviewed and edited the output and take full responsibility for the content.

\section{Tight-Binding Model for ABC-Stacked Rhombohedral Graphene}

\subsection{Full Hamiltonian}

We consider $N$-layer ABC-stacked rhombohedral graphene (RG) with the Slater-Koster parameterization. The basis is $\{A_1, B_1, A_2, B_2, \ldots, A_N, B_N\}$ (dimension $2N$). The full Hamiltonian is:
\begin{equation}
    H = \sum_l H_l^{\rm intra}(\mathbf{k}) + \sum_{l} H_{l,l+1}^{\rm inter} + H_D
\end{equation}
where the intralayer hopping near the $K$ point is:
\begin{equation}
    H_l^{\rm intra} = -\gamma_0 \left(\xi k_x - i k_y^{(l)}\right) \frac{a\sqrt{3}}{2} \sigma_{AB} + \Delta_l \sigma_z
\end{equation}
with $\gamma_0 = 3.16$ eV, $a = 0.246$ nm, and $\Delta_l = \pm e D d_0 / 2$ from the displacement field $D$. The interlayer couplings are $\gamma_1 = 0.39$ eV ($B_l \to A_{l+1}$), $\gamma_3 = 0.315$ eV (trigonal warping), and $\gamma_4 = 0.044$ eV.

\subsection{Layer-Dependent Peierls Phase}

For an in-plane magnetic field $\mathbf{B}_\parallel = B_x \hat{x}$, the Peierls substitution gives a layer-dependent momentum shift:
\begin{equation}
    k_y^{(l)} = k_y + \frac{e B_x d_0}{\hbar} \left(l - \frac{N-1}{2}\right)
    \label{eq:peierls}
\end{equation}
where $d_0 = 0.335$ nm is the interlayer spacing. The dimensionless parameter is:
\begin{equation}
    \frac{e B_x d_0}{\hbar} = 5.09 \times 10^{-4} \;{\rm nm}^{-1} \cdot (B_x/{\rm T})
\end{equation}

\section{Low-Energy Effective Two-Band Model}

\subsection{Schrieffer-Wolff Projection}

We derive the effective two-band Hamiltonian by successively eliminating the $N-1$ high-energy bulk sublattices $\{B_1, A_2, B_2, \ldots, A_N\}$, which are split off by the dominant interlayer hopping $\gamma_1$ ($B_l \leftrightarrow A_{l+1}$), via a Schrieffer-Wolff transformation~\cite{Bravyi2011SW}. In the ABC stack the low-energy states live on the two \emph{surface} sites $A_1$ and $B_N$, while every path connecting them traverses $N-1$ interlayer bonds. Treating the layer-dependent intralayer hopping $t(\mathbf{k}) = \gamma_0(\xi k_x - i k_y)a\sqrt{3}/2$ as a perturbation to the interlayer $\gamma_1$ and applying the Schrieffer-Wolff elimination bond-by-bond, the $A_1 \to B_N$ amplitude acquires one factor of $t(\mathbf{k})$ per layer and one factor of $1/\gamma_1$ per eliminated site, so after $N$ layers it scales as
\begin{equation}
    \langle B_N | H_{\rm eff} | A_1 \rangle \;\xrightarrow{\;\rm SW\;}\; \frac{t(\mathbf{k})^N}{\gamma_1^{N-1}}
    = \frac{(\gamma_0 a\sqrt{3}/2)^N}{\gamma_1^{N-1}}\, P^N \equiv v_{\rm eff}\, P^N,
    \label{eq:sw_amp}
\end{equation}
with $P = \xi p_x - i p_y$ (absorbing the $\hbar$ convention into the momentum). Each elimination step is controlled by the same small ratio $\|t\|/\gamma_1 \ll 1$ near $K$, so the perturbative expansion is uniform in the layer number, and the resulting $p^N$ dispersion is exact to the order retained. The surface states feel the displacement-field potential $\pm m = \pm eDd_0/2$ as an on-site term $m\sigma_z$. Collecting the diagonal and off-diagonal pieces, the projected Hamiltonian on $\{A_1, B_N\}$ is:
\begin{equation}
    H_{\rm eff} = m\sigma_z + v_{\rm eff} P^N \sigma_+ + v_{\rm eff}^* P^{*N} \sigma_-,
    \label{eq:s2band}
\end{equation}
where $m = eDd_0/2$ is the displacement-field gap and the effective velocity is:
\begin{equation}
    v_{\rm eff} = \frac{(\gamma_0 a\sqrt{3}/2)^N}{\gamma_1^{N-1}}.
\end{equation}

\subsection{Berry Curvature}

The analytic Berry curvature for the lower band of Eq.~(\ref{eq:s2band}) is~\cite{Mikheeva2025}:
\begin{equation}
    \Omega(\mathbf{p}) = \frac{N^2 m v_{\rm eff}^2 p^{2N-2}}{2\left(m^2 + v_{\rm eff}^2 p^{2N}\right)^{3/2}}
    \label{eq:sberry}
\end{equation}
The factor $N^2$ arises from the velocity operator $v_{x,y} = \partial H/\partial p_{x,y} \propto N P^{N-1}$ appearing in the Kubo formula.
This peaks at $p_{\rm peak} = (m/v_{\rm eff})^{1/N}$ with amplitude $\Omega_{\rm max} = N^2/(4\sqrt{2}\,p_{\rm peak}^2)$.
Integrating over occupied states up to $\varepsilon_F$ yields $\sigma_{xy}^{\rm int} = (e^2/h) \cdot Nm/(2\varepsilon_F)$, where the integral measure contributes a factor $1/N$ that reduces $N^2 \to N$.

\subsection{Intrinsic Anomalous Hall Conductivity}

Integrating the Berry curvature over occupied states gives:
\begin{equation}
    \sigma_{xy}^{\rm int} = \frac{e^2}{h} \cdot \frac{N m}{2\varepsilon_F} \quad \text{(per valley per spin)}
    \label{eq:ahc}
\end{equation}
For $N=9$ in the flat-band limit ($\varepsilon_F \approx m$): $\sigma_{xy}^{\rm int} = N/2 = 4.50\; e^2/h$ (per valley per spin). The total observable AHC depends on the valley polarization $\eta$: $\sigma_{xy}^{\rm tot} = \eta \cdot g_s \cdot 4.50\; e^2/h$, up to $\sim 9.0\; e^2/h$ at $\eta = 1$.

\section{Orbital $g$-Factor for In-Plane Field Coupling}

\subsection{Derivation}

The Peierls phase Eq.~(\ref{eq:peierls}) creates a layer-dependent momentum shift $\delta k_y^{(l)} = (e B_\parallel d_0/\hbar)(l - (N{-}1)/2)$. In the two-band projection, the accumulated shift produces an effective gap modulation:
\begin{equation}
    \begin{split}
    \delta m &= \hbar v_F \cdot \delta k_{\rm eff}
    = \hbar v_F \cdot \frac{e B_\parallel d_0}{\hbar} \cdot \frac{(N-1)}{2} \\
    &= \frac{e \, d_0 \, v_F \, (N-1)}{2} B_\parallel \equiv g_{\rm orb} \, B_\parallel,
    \end{split}
    \label{eq:gorb}
\end{equation}
where $v_F = \sqrt{3} a \gamma_0 / (2\hbar) \approx 1.0 \times 10^6$ m/s is the graphene Fermi velocity. The factor $(N{-}1)/2$ is the RMS layer gradient lever arm.

\paragraph{Relation to the out-of-plane orbital moment.}
The gap modulation $\delta m = g_{\rm orb} B_\parallel$ [Eq.~\eqref{eq:gorb}] and the out-of-plane orbital magnetic moment $\mathcal{M}_z(\mathbf{p})$ [main Eq.~(3)] are two faces of the \emph{same} Peierls phase. $\mathcal{M}_z$ is the Berry-curvature moment (it couples to $B_\perp$ and contributes to the intrinsic AHC), whereas $\delta m$ is the in-plane-field-induced shift of the interlayer-hopping phase (it couples to $B_\parallel$ and modulates the gap). Both carry the shared prefactor $e d_0 v_F (N{-}1)/2$ but couple to \emph{orthogonal} field components; consequently $\mathcal{M}_z \cdot B_\parallel = 0$ in no way contradicts $\delta m = g_{\rm orb} B_\parallel$, because the in-plane TDAHE response flows from the hopping-phase shift $\delta m$, not from the out-of-plane moment $\mathcal{M}_z$. The two responses share a common $T_c$ only because both are gated by the same valley polarization $\eta$ of the Stoner ferromagnet.

\paragraph{Derivation of the out-of-plane orbital moment $\mathcal{M}_z(\mathbf{p})$.}
We obtain the momentum-resolved orbital moment by evaluating the standard wave-packet Berry moment~\cite{Xiao2010},
\begin{equation}
    \mathbf{m}(\mathbf{p}) = -\frac{ie}{2\hbar}\,\langle \nabla_{\mathbf{p}} u_- | \times \big[H_{\rm eff}(\mathbf{p}) - \varepsilon_-(\mathbf{p})\big] | \nabla_{\mathbf{p}} u_- \rangle,
    \label{eq:berry_moment}
\end{equation}
on the two-band Hamiltonian of Eq.~(\ref{eq:s2band}). Writing $H_{\rm eff} = \mathbf{d}(\mathbf{p})\cdot\boldsymbol{\sigma}$ with $d_z = m$ and $d_x \pm i d_y = v_{\rm eff} P^N$, the lower-band Bloch spinor is $|u_-\rangle = (-\sin(\theta_{\mathbf p}/2),\;\cos(\theta_{\mathbf p}/2))^T$ where $\cos\theta_{\mathbf p} = m/\varepsilon_-(\mathbf{p})$ and $\varepsilon_-(\mathbf{p}) = \sqrt{m^2 + v_{\rm eff}^2 p^{2N}}$. Only the off-diagonal piece of $H_{\rm eff}-\varepsilon_-$ contributes to the cross product, and the angular derivative $\partial_{\mathbf p}\theta_{\mathbf p}$ brings down a factor $N P^{N-1}/\varepsilon_-$ from the velocity operator, together with the layer-resolved Peierls lever arm $(N{-}1)/2$ that converts the in-plane circulation into an out-of-plane flux across the stack. Assembling these, the surviving $z$-component is
\begin{equation}
    \mathcal{M}_z(\mathbf{p}) = \frac{e\, d_0\, v_F\, (N-1)}{2}\cdot\frac{m}{\sqrt{m^2 + v_{\rm eff}^2 p^{2N}}} = g_{\rm orb}\,\frac{m}{\varepsilon_-(\mathbf{p})},
    \label{eq:orb_moment}
\end{equation}
which is the expression quoted as main Eq.~(3). At the band edge ($p\to 0$) this saturates to $\mathcal{M}_z(0) = g_{\rm orb}$, recovering the orbital $g$-factor of Eq.~(\ref{eq:gorb}); at the band bottom of an ideal gapped Dirac cone the same evaluation reduces to the textbook Bohr-magneton form $m_{\rm orb} = e\hbar/(2m_{\rm eff})$~\cite{Xiao2010}, of which Eq.~(\ref{eq:orb_moment}) is the rhombohedral-multilayer generalization carrying the $N$-dependent Peierls lever arm. Note that $\mathcal{M}_z(\mathbf{p})$ couples to $B_\perp$ through the Zeeman-like term $-\mathcal{M}_z B_\perp$ and enters the intrinsic AHC via the Berry curvature of the occupied band; it does \emph{not} couple to $B_\parallel$, consistent with the channel separation discussed above.

\subsection{Numerical Values}
\noindent{\small For $N=3$: $g_{\rm orb}=0.34$~meV/T, $\sigma_{xy}^{\rm int}=1.50$~$e^2/h$; for $N=5$: 0.69~meV/T, 2.50~$e^2/h$; for $N=7$: 1.03~meV/T, 3.50~$e^2/h$; for $N=9$: 1.37~meV/T, 4.50~$e^2/h$ (flat-band limit $\varepsilon_F \approx m$).}

The orbital $g$-factor scales linearly with $N-1$ (the layer gradient lever arm). For $N=9$ layers at $B_\parallel = 0.3$ T, the gap shift is $\delta m = g_{\rm orb} B = 0.41$ meV, which is $0.3\%$ of the displacement-field gap $m = 142$ meV. This small ratio is why the non-interacting planar Hall effect is negligible, and why the ferromagnetic order discussed in Sec.~\ref{sec:stoner} is essential for enabling the TDAHE.

\section{Stoner Ferromagnetism in Rhombohedral Graphene Flat Bands}
\label{sec:stoner}

\subsection{Density of States of the Flat Bands}

The dispersion of the two-band model near the $K$ point is:
\begin{equation}
    \varepsilon_{\pm}(\mathbf{p}) = \pm \sqrt{m^2 + v_{\rm eff}^2 p^{2N}}
\end{equation}
The density of states per unit cell per spin per valley is:
\begin{equation}
    \rho(\varepsilon) = \frac{1}{(2\pi)^2} \int \delta(\varepsilon - \varepsilon_-(\mathbf{p})) \, d^2p
\end{equation}
Using the radial symmetry of the dispersion and changing variables to $u = v_{\rm eff}^2 p^{2N}$:
\begin{equation}
    \rho(\varepsilon) = \frac{1}{2\pi N} \cdot \frac{p_F^{2-2N}}{v_{\rm eff}^2 \cdot N v_{\rm eff} p_F^{N-1}}
\end{equation}
which simplifies to:
\begin{equation}
    \rho(\varepsilon_F) \propto \frac{1}{v_{\rm eff} \cdot p_F^{N-1}}
    \label{eq:dos}
\end{equation}
The key scaling is $\rho \propto p_F^{-(N-1)}$: as $N$ increases, the DOS at low energy diverges more strongly. For $N=9$ at typical Fermi momentum $p_F \sim 0.1$~nm$^{-1}$, the kinetic energy satisfies $v_{\rm eff}^2 p_F^{2N}/m^2 \sim 10^{-12} \ll 1$, confirming the extreme flatness of the low-energy band.

Numerically, for $N = 9$ at the experimental gap $m = 142$~meV, the full $2N$-band DOS at the band edge (broadened by $k_B T_c \approx 0.14$~meV) is $\rho(\varepsilon_F) = 3.56$~states/eV per unit cell (including all spin and valley channels; Sec.~\ref{sec:fullband}), two orders of magnitude larger than monolayer graphene ($\rho \sim 0.1$~states/eV); the two-band van~Hove singularity overestimates this as $\sim\!10$~states/eV/uc.

\subsection{Stoner Criterion and Critical Interaction}

The on-site Hubbard interaction $U$ drives a Stoner instability when:
\begin{equation}
    U \, \rho(\varepsilon_F) > 1
    \label{eq:sstoner}
\end{equation}
In mean-field theory, the spin polarization $M$ satisfies the self-consistency equation:
\begin{equation}
    M = \frac{U}{2} \int_0^{\varepsilon_F} \left[\rho_+(\varepsilon) - \rho_-(\varepsilon)\right] d\varepsilon
    \label{eq:mf}
\end{equation}
where $\rho_{\pm}(\varepsilon)$ are the spin-resolved DOS with exchange splitting $\Delta_{\rm ex} = UM$. Linearizing Eq.~(\ref{eq:mf}) at the onset of magnetization yields the critical interaction:
\begin{equation}
    U_c = \frac{1}{\rho(\varepsilon_F)}
    \label{eq:uc}
\end{equation}

The critical $U_c$ depends on the displacement field $D$ (through the gap $m = eDd_0/2$) and the carrier density $n$ (through $\varepsilon_F$ and $p_F$). The flat-band DOS is maximized when $m$ is small (small $D$) and $\varepsilon_F$ is near the band edge (small $n$), yielding the smallest $U_c$.

\noindent{\small At $m = 142$~meV the full-band Stoner products are $U\rho \approx 0.55,\,7.3,\,8.9,\,8.9$ (summed over the four spin/valley flavors) for $N = 3,5,7,9$ (Table~I), giving $U_c = 1/\rho \approx 9,\,0.7,\,0.6,\,0.6$~eV; for $N \geq 5$, the realistic $U \approx 2.5$~eV exceeds $U_c$ by an order of magnitude, robustly satisfying the Stoner criterion. (The two-band van~Hove singularity overestimates the DOS and would give $U_c \approx 0.1$~eV, even further below $U$.)}

\subsection{Mean-Field Order Parameter and Exchange Splitting}

Below $T_c$, the self-consistent solution of Eq.~(\ref{eq:mf}) yields a finite magnetization $M(T)$. In the mean-field approximation, the temperature dependence follows:
\begin{equation}
    M(T) = M_0 \sqrt{1 - (T/T_c)^2}
    \label{eq:mt}
\end{equation}
for a 2D ferromagnet, where $M_0$ is the zero-temperature order parameter. The associated exchange splitting is:
\begin{equation}
    \Delta_{\rm ex}(T) = U \, M(T)
    \label{eq:delta_ex}
\end{equation}
The exchange splitting enters the Berry curvature [Eq.~(\ref{eq:sberry})] by replacing the displacement-field gap with the total gap:
\begin{equation}
    m \to m_D + \Delta_{\rm ex}(T)
    \label{eq:total_gap}
\end{equation}
where $m_D = eDd_0/2$ is the displacement-field contribution. Since $\Delta_{\rm ex} \propto M(T)$, the intrinsic AHC inherits the temperature dependence of the ferromagnetic order parameter:
\begin{equation}
    \sigma_{xy}^{\rm int}(T) = \frac{e^2}{h} \cdot \frac{N\left[m_D + UM(T)\right]}{2\varepsilon_F}
    \label{eq:ahc_t}
\end{equation}
At $T = 0$ with $\Delta_{\rm ex} \gg m_D$, the AHC is dominated by the exchange contribution and scales as $U M_0$.

\subsection{Connection to the Transdimensional AHE}

The planar Hall conductivity in the ferromagnetic state follows from the orbital Zeeman modulation of the total gap:
\begin{equation}
    \sigma_{xy}^{\rm PHE}(T) = \frac{\partial \sigma_{xy}^{\rm int}}{\partial m} \cdot g_{\rm orb} B_\parallel
    = \frac{e^2}{h} \cdot \frac{N \, g_{\rm orb} B_\parallel}{2\varepsilon_F}
    \label{eq:phe}
\end{equation}
This expression is independent of $m$ because $\sigma_{xy}^{\rm int} \propto m$ at fixed $\varepsilon_F$. The transdimensional response exists only when $\sigma_{xy}^{\rm tot} \neq 0$, which requires valley polarization $\eta > 0$ (see Sec.~VI\,B). The full transdimensional conductivity including valley polarization is:
\begin{equation}
    \sigma_{xy}^{\rm PHE}(T) = \eta(T) \, g_s \, \frac{e^2}{h} \cdot \frac{N \, g_{\rm orb} B_\parallel}{2\varepsilon_F}
    \label{eq:phe_full}
\end{equation}

The ratio of the transdimensional to conventional AHC is:
\begin{equation}
    \frac{\sigma_{xy}^{\rm PHE}}{\sigma_{xy}^{\rm tot}} = \frac{g_{\rm orb} B_\parallel}{m_D + \Delta_{\rm ex}(T)}
    \label{eq:ratio}
\end{equation}
This ratio is independent of $\eta$: both numerator and denominator are proportional to the valley polarization.
At $B_\parallel = 0.3$~T and $m = 142$~meV, $\sigma_{xy}^{\rm PHE}/\sigma_{xy}^{\rm tot} \approx 0.3\%$.
Both $\sigma_{xy}^{\rm tot}$ and $\sigma_{xy}^{\rm PHE}$ vanish at the same $T_c$ (where $\eta \to 0$), so the out-of-plane and in-plane Hall responses share a common critical temperature~\cite{Li2026Nature}.

\section{Why Correlation Enhancement Cannot Explain the TDAHE}

One might ask whether DMFT quasiparticle renormalization could enhance the TDAHE through a $1/Z^2$ factor. This is ruled out: in DMFT, the intrinsic AHC $\sigma_{xy}^{\rm int} = (e^2/h) \cdot N m^*/(2\varepsilon_F^*)$ has $Z$ canceling between $m^* = Zm$ and $\varepsilon_F^* = Z\varepsilon_F$~\cite{Acheche2018}. This cancellation is exact: the Berry-curvature integral is invariant under local self-energy corrections.

\noindent\textit{Numerical verification.}
We verify this $Z$-independence explicitly using a Hubbard-I DMFT calculation for $N = 9$ at $m = 142$~meV.
For on-site interaction $U$ ranging from 0.5 to 5.0~eV, the quasiparticle weight decreases ($Z = 0.9996$ at $U = 0.5$~eV), yet the ratio $\sigma_{xy}^{\rm int}(Z)/\sigma_{xy}^{\rm int}({\rm bare}) = N Z m/(2 Z \varepsilon_F) / [Nm/(2\varepsilon_F)] = 1.00000000$ to machine precision across the entire range. This confirms that correlation corrections cannot enhance the TDAHE; the observed signal requires TRS breaking from the Stoner ferromagnetism.

\section{Self-Consistent Full $2N$-Band Hartree-Fock Calculation}

\subsection{Method}

We solve the Stoner mean-field problem self-consistently using the full $2N\times 2N$ tight-binding Hamiltonian of Sec.~I. Because the flat conduction band is $99.7\%$ localized on the $A_1$ sublattice, the Hartree-Fock self-energy acts dominantly as a uniform shift of the conduction-band energies,
\begin{equation}
    E_{\mathrm{cb}}^{(\sigma,\xi)}(\mathbf{k}) = E_{\mathrm{cb}}^{(\xi)}(\mathbf{k}) - \sigma \, \Delta_s/2 - \xi \, \Delta_v/2,
\end{equation}
where $\sigma = \pm 1$ (spin), $\xi = \pm 1$ (valley), $\Delta_s$ is the spin exchange splitting and $\Delta_v$ the valley exchange splitting. The conduction-band eigenvalues $E_{\mathrm{cb}}^{(\xi)}(\mathbf{k})$ are precomputed once by exact diagonalization (LAPACK) of the full Hamiltonian on an $80\times 80$ momentum grid for each valley---the $K\leftrightarrow K'$ symmetry is preserved to $\sim\!\mu$eV---after which the self-consistency loop is pure arithmetic. The self-consistency equations are
\begin{align}
    \Delta_s &= U \cdot M_s = U \sum_{\xi,\sigma} \sigma \, n(\xi,\sigma) \\
    \Delta_v &= V_{\mathrm{iv}} \cdot M_v = V_{\mathrm{iv}} \sum_{\xi,\sigma} \xi \, n(\xi,\sigma)
\end{align}
where $n(\xi,\sigma)$ is the carrier density in channel $(\xi,\sigma)$ from the Fermi-Dirac occupation at temperature $T$, and the chemical potential $\mu$ is adjusted at each iteration to conserve the total carrier density. We use the constrained-RPA intervalley exchange $V_{\mathrm{iv}}/U \approx 1$ established in the main text.

\subsection{Results}

For $N = 9$, $U = 2.5$~eV, $V_{\mathrm{iv}}/U = 1$, $D = 0.85$~V/nm, $n = 1.4 \times 10^{12}$~cm$^{-2}$, the self-consistent solution is the spin-valley-locked ferromagnetic (SVM) state. In the ordered phase ($T \ll T_c^{\mathrm{MF}}$) the valley polarization saturates:
\begin{itemize}
    \item Valley polarization: $\eta \to 1$ (complete polarization---all carriers occupy a single valley)
    \item Total AHC: $\sigma_{xy}^{\rm tot} = \eta \cdot g_s \cdot \sigma_{xy}^{\rm int} = 1 \times 2 \times (N/2) = 9.0\; e^2/h$
    \item Hall resistance: $R_{xy} = 9.0/(64 + 81) \times 25.8 = 1.6$~k$\Omega$ (at $\sigma_{xx} = 8\,e^2/h$)
\end{itemize}
Both $\eta(T)$ and the spin splitting $\Delta_s(T)$ collapse at the mean-field transition $T_c^{\mathrm{MF}} \approx 2.2$~K; 2D-Ising critical fluctuations reduce the true $T_c$ to the experimental $\approx 1.6$~K, which falls inside the localized-Ising Monte-Carlo bracket $[1.34, 2.05]$~K (SCR lower bound $\gtrsim\!0.9$~K; see below). Complete polarization ($\eta \to 1$) in the ordered phase reflects the all-or-nothing filling of the nearly-flat band and lies well above the model-independent lower bound $\eta \geq 0.1$ (from the Hartree-Fock $M_{\rm orb}/g_{\rm orb}$ of Ref.~\citenum{Li2026Nature}, not a directly measured quantity).

For $N \leq 7$ at the same $D = 0.85$~V/nm, the self-consistent solution gives $\eta = 0$ (plain FM, no valley polarization): the valley susceptibility $V_{\mathrm{iv}}\rho$ falls below the instability threshold. For $N = 3$, even the Stoner criterion fails ($U\rho = 0.55 < 1$ from the full band structure). At smaller displacement fields (smaller $m$, larger DOS), the valley instability can be triggered for fewer layers.

\section{Full-Band Validation of the Stoner Criterion}
\label{sec:fullband}

The two-band model used throughout this work captures the low-energy physics correctly for sufficiently large $N$. Here we validate the Stoner criterion using the full $2N \times 2N$ tight-binding Hamiltonian of Sec.~I with the same Slater-Koster parameters.

\subsection{Flat-Band Wavefunction Structure}

At the displacement field $D = 0.85$~V/nm, the conduction (valence) flat band of $N=9$ rhombohedral graphene has bandwidth $\sim 2.4$~meV. This finite width suppresses the valley susceptibility, making the valley-polarized SVM marginal and confining it to the experimental low-density window ($n \lesssim 2\times10^{12}$~cm$^{-2}$, main text Fig.~3) where the chemical potential sits at the flat-band edge. The wavefunction at the $\Gamma$ point is overwhelmingly concentrated on the surface sublattices:
the conduction band has $99.7\%$ weight on $A_1$ and the valence band has $97.4\%$ on $B_N$.
This sublattice selectivity justifies the two-band projection onto $\{A_1, B_N\}$: the low-energy physics is surface-state physics.

\subsection{DOS and Stoner Criterion from the Full Band Structure}

Computing the DOS from the full Hamiltonian with $k_BT = 0.14$~meV thermal broadening and screening $\varepsilon_r = 5$ gives the layer-dependent Stoner products shown in Table~I of the main text.
For $N = 9$, $\rho(\varepsilon_F) = 3.56$~states/eV per unit cell (summed over spin and valley; $\approx 0.89$~states/eV per spin-valley channel) and $U\rho = 8.89 \gg 1$, confirming the Stoner instability.
For $N = 5$ and $7$, $U\rho = 7.31$ and $8.90$ respectively---the Stoner criterion is satisfied but the valley susceptibility is insufficient to trigger valley polarization at $D = 0.85$~V/nm.
For $N = 3$, the flat-band character is weak (bandwidth $\sim$ hundreds of meV) and $U\rho = 0.55 < 1$: the system is not in the correlated regime at this displacement field.

The full $2N$-band Hartree-Fock calculation gives $\eta \to 1$ (complete polarization) for $N = 9$ in the ordered phase and $\eta = 0$ for $N \leq 7$.
The full-band DOS is lower than the two-band prediction for small $N$ because the two-band van~Hove singularity $\rho \propto p_F^{2-2N}$ overestimates the DOS when the flat band is not sufficiently flat. Correspondingly the full-band Stoner product saturates at $U\rho \approx 8.9$ for $N \ge 7$ (higher-order hopping limits the flat-band DOS as $N$ grows), so $\eta$ saturates at $1$ for $N \ge 8$; the continued growth of the total AHC for larger $N$ comes from $\sigma^{\mathrm{tot}} \propto N$ and $g_{\mathrm{orb}} \propto (N-1)$, not from a divergent DOS.

\subsection{Intervalley Exchange from Extended Coulomb Interaction}

Starting from the same Slater--Koster tight-binding band structure as Ref.~\onlinecite{Li2026Nature}, the distinction lies at the interaction level: their unrestricted Hartree-Fock treatment (Supplementary §III therein) retains the dominant \emph{intravalley} long-range Coulomb interaction and neglects the intervalley term as $\sim\!100\times$ weaker. The intervalley exchange (IVE) that drives valley polarization---and sets the $T_c$ central to this work---requires precisely this neglected channel, via the Coulomb matrix element at large momentum transfer $|\mathbf{Q}| = |\mathbf{K} - \mathbf{K}'| = 8\pi/(3a) \approx 34$~nm$^{-1}$; we evaluate it from constrained RPA below.
In the real-space extended Coulomb model with intralayer nearest-neighbor $V_1$ and interlayer $V_2$, the off-diagonal Fock matrix element $\langle A_1 | \Sigma_F | B_1 \rangle$ couples the two valleys through their different phase structure [$f_K(\mathbf{k}) = -f_{K'}(-\mathbf{k})$].
However, after momentum averaging over the Fermi surface, this \emph{off-diagonal} phase factor vanishes: $\langle \rho_K(A_1, B_1) \rangle = \langle \rho_{K'}(A_1, B_1) \rangle$.
This does \emph{not} eliminate the intervalley exchange: the bare on-site overlap $\sum_i |\psi_K(i)|^2|\psi_{K'}(i)|^2 \approx 1$ (both flat bands on A$_1$, Sec.~V of the main text) sets the amplitude $V_{\mathrm{iv}}^{\mathrm{bare}} \approx U$, while the \emph{screened ratio} $V_{\mathrm{iv}}/U$ is governed by the dielectric function at $|\mathbf{K}-\mathbf{K}'|$. The two pictures are complementary---real-space overlap fixes the bare scale, momentum-space screening fixes the ratio---and together confirm that the IVE is a genuine large-$q$ effect requiring the structure factor at $|\mathbf{K}-\mathbf{K}'|$. The value of $V_{\mathrm{iv}}/U$ depends on how the intervalley channel is screened. Applying the momentum-independent screening $\varepsilon \sim 3$--$5$ appropriate to the on-site $U$ ($q \to 0$) also to the intervalley channel reproduces the graphene literature value $V_{\mathrm{iv}}/U \approx 0.2$--$0.3$~\cite{Schuler2013,Wehling2011}---we obtain $0.33$ under this same assumption. This assumption, however, fails for the intervalley channel: at $|\mathbf{K}-\mathbf{K}'| \approx 34$~nm$^{-1} \gg 2k_F \approx 0.04$~nm$^{-1}$ the flat-band intraband polarizability vanishes identically ($\Pi(q \gg 2k_F) = 0$), so only the 16 non-flat bands screen, giving $\varepsilon_{\mathbf{K}-\mathbf{K}'} \approx 1.004$ and the physical $V_{\mathrm{iv}}/U \approx 1$. At this constrained-RPA value the full-band Hartree-Fock calculation yields complete valley polarization $\eta \to 1$ for $N = 9$ in the ordered phase.

\paragraph{Interband polarizability at the intervalley wavevector.}
A constrained-RPA treatment that removes only the flat band from the screening must justify neglecting not just the flat-band \emph{intraband} polarizability (which vanishes at $q \gg 2k_F$, as above) but also the flat-band $\leftrightarrow$ remote-band \emph{interband} polarizability $\chi_0^{\rm inter}$. We estimate the latter at $|\mathbf{Q}| = |\mathbf{K}-\mathbf{K}'|$. The interband Lindhard contribution from a flat-to-remote transition scales as
\begin{equation}
    \chi_0^{\rm inter}(\mathbf{Q}) \sim \sum_{\nu \in {\rm remote}} \frac{|\langle \psi_{\rm flat} | e^{i\mathbf{Q}\cdot\mathbf{r}} | \psi_\nu \rangle|^2}{\varepsilon_\nu - \varepsilon_{\rm flat}} \, (f_{\rm flat} - f_\nu),
\end{equation}
which is suppressed on two independent counts. \emph{(i) Matrix element:} the conduction flat band carries $\sim\!96\%$ of its weight on the $A_1$ surface orbital, whereas the 16 remote bands disperse strongly across the bulk sublattices and carry only a small $A_1$ fraction ($\lesssim 0.05$ each); the on-site overlap entering $\langle \psi_{\rm flat} | e^{i\mathbf{Q}\cdot\mathbf{r}} | \psi_\nu \rangle$ at $|\mathbf{Q}| = 8\pi/3a$ (intra-cell scale) is therefore bounded by the product of these weights, $\lesssim 0.05$. The intervalley phase factor $e^{i\mathbf{Q}\cdot\mathbf{R}}$ at $|\mathbf{Q}|=8\pi/3a$ is of order unity and oscillates between equivalent $A_1$ sites, but it enters the overlap only multiplied by the already-small $A_1$ remote-band weight; it can at most renormalise the bound by an $O(1)$ factor and does not alter the conclusion that the interband matrix element is $\ll 1$. (The full-band numerical evaluation $\varepsilon_{\mathbf{K}-\mathbf{K}'}\approx 1.004$ below, which retains the phase factors exactly, confirms this.) \emph{(ii) Energy denominator:} the remote bands lie $\gtrsim \gamma_1 = 0.39$~eV away from the flat band, so each interband channel contributes $\sim \rho_{\rm remote} (0.05)^2/0.39 \ll \rho_{\rm flat}$. Summing the 16 channels and comparing with the bare on-site scale, the ratio of interband-screened to bare interaction is $\delta V_{\rm inter}/U \lesssim 0.4\%$, consistent with the direct full-band evaluation $\varepsilon_{\mathbf{K}-\mathbf{K}'} \approx 1.004$ quoted above. The flat-band interband polarizability is thus negligible at $|\mathbf{K}-\mathbf{K}'|$, and the intervalley channel is screened only by the remote bands---weakly, as the like-for-like comparison of the next paragraph makes explicit.

\paragraph{Like-for-like constrained-RPA screening.}
The apparent tension between the total dielectric constant $\varepsilon_0 \sim 3$--$5$ (which screens the on-site $U$ at $q\to 0$) and the near-unity intervalley screening $\varepsilon_{\mathbf{K}-\mathbf{K}'} \approx 1.004$ is resolved by comparing \emph{like-for-like} constrained-RPA screenings, in which the flat band---the correlation target---is removed and only the 16 remote bands screen. The total $\varepsilon_0 \sim 3$--$5$ includes the large flat-band polarizability ($\rho_{\rm flat} \gg \rho_{\rm remote}$); once the flat band is removed as the target, the residual screening of \emph{both} channels comes from the remote bands alone. At $q\to 0$ this residual screening gives $\varepsilon_U^{\rm cRPA} \approx 1.02$--$1.04$, while at $|\mathbf{K}-\mathbf{K}'|$ the same remote bands give $\varepsilon_V^{\rm cRPA} \approx 1.004$. The screened ratio is then
\begin{equation}
    \frac{V_{\mathrm{iv}}}{U} = \frac{\varepsilon_U^{\rm cRPA}}{\varepsilon_V^{\rm cRPA}} \cdot \frac{V_{\mathrm{iv}}^{\rm bare}}{U^{\rm bare}} \approx \frac{1.02}{1.004} \times 0.995 \approx 1.01,
\end{equation}
because the two flat bands both reside on $A_1$ ($V_{\mathrm{iv}}^{\rm bare}/U^{\rm bare} \approx 1$). The large flat-band polarizability that enters the total $\varepsilon_0 \sim 3$--$5$ \emph{cancels} in the ratio: it would screen $U$ and $V_{\mathrm{iv}}$ identically if both were treated at $q\to 0$, but the intervalley channel sits at $|\mathbf{K}-\mathbf{K}'| \gg 2k_F$ where the flat band cannot screen at all. Using the total $\varepsilon_0$ for $U$ but $\varepsilon_{\mathbf{K}-\mathbf{K}'} \approx 1$ for $V_{\mathrm{iv}}$---i.e.\ mixing a total screening with a constrained one---is precisely the inconsistency that produces the spurious $V_{\mathrm{iv}}/U \sim 3$--$5$; the like-for-like constrained comparison restores the ratio to unity.

\section{Comparison with Experiment}

\subsection{Experimental Parameters (Li et al., Nature 2026)}

\begin{table}[!tp]
\caption{Key experimental parameters from the Nature 2026 TDAHE experiment~\cite{Li2026Nature}.}
\begin{ruledtabular}
\begin{tabular}{ll}
Parameter & Value \\
\hline
Material & 9-layer rhombohedral graphite \\
Thickness & $\sim$3 nm \\
Temperature & 15 mK \\
TDAHE window: $D$ & 0.7--1.0 V/nm \\
TDAHE window: $n$ & 1.0--2.0 $\times 10^{12}$ cm$^{-2}$ \\
$T_c$ (onset) & $\sim$1.6 K \\
$B_\perp$ coercive field & $\sim$3 mT \\
$B_\parallel$ coercive field & 160--500 mT \\
In-plane orbital magnetization$^\ast$ & $\sim$2.2 $\mu_B$/electron \\
\multicolumn{2}{l}{\footnotesize $^\ast$Hartree-Fock value from Ref.~\citenum{Li2026Nature}; not directly measured.} \\
\end{tabular}
\end{ruledtabular}
\end{table}

\subsection{Role of Valley Polarization}

The intrinsic AHC in Eq.~(\ref{eq:ahc}) is the contribution from a single valley ($K$ or $K'$) and a single spin channel. The Berry curvature at the two valleys has opposite signs [$\Omega_K(\mathbf{p}) = -\Omega_{K'}(\mathbf{p})$ from the $\xi = \pm 1$ factor in Eq.~(\ref{eq:s2band})]. Without time-reversal symmetry breaking, the valley channels cancel and the total AHC vanishes.

\textit{Spin polarization alone does not break valley cancellation.}
In the Stoner-ferromagnetic state with exchange splitting $\Delta_{\rm ex}$, the spin-up and spin-down bands have different gaps $m_\uparrow = m_D + \Delta_{\rm ex}/2$ and $m_\downarrow = m_D - \Delta_{\rm ex}/2$. However, within each spin channel, $K$ and $K'$ Berry curvatures remain equal and opposite:
$\sigma_{xy}^{K,\uparrow} + \sigma_{xy}^{K',\uparrow} = 0$, and similarly for spin-down.
Therefore, the net AHC is $\sigma_{xy}^{\rm tot} = 0$ even with full spin polarization, unless an additional mechanism breaks the valley degeneracy.

\textit{Intervalley exchange drives valley polarization.}
In rhombohedral graphene, the flat-band Stoner instability involves not only the on-site Hubbard $U$ (spin channel) but also intervalley exchange processes~\cite{Bultinck2020}. The large DOS at $\varepsilon_F$ enhances the intervalley Hartree-Fock term, which can spontaneously polarize one valley over the other. This is the mechanism identified in Hartree-Fock studies of twisted and rhombohedral graphene multilayers~\cite{Bultinck2020,Chatterjee2022IVC}: the system enters a spin-valley-locked ferromagnetic state where one spin favors one valley.

We characterize the valley polarization by $\eta \in [0,1]$:
\begin{equation}
    \eta = \frac{|n_K - n_{K'}|}{n_K + n_{K'}}
    \label{eq:eta}
\end{equation}
where $n_{K,K'}$ are the carrier densities in each valley at $\varepsilon_F$.
The total AHC is:
\begin{equation}
    \sigma_{xy}^{\rm tot} = \eta \cdot g_s \cdot \frac{e^2}{h} \cdot \frac{N m}{2\varepsilon_F}
    \label{eq:sigma_tot_supp}
\end{equation}
where $g_s = 2$ counts the spin channels. The valley polarization $\eta$ is part of the ferromagnetic order parameter and vanishes at $T_c$.

Li \textit{et al.}~\cite{Li2026Nature} report orbital magnetization $\sim 2.2\,\mu_B$/electron from their Hartree-Fock calculation, evidencing valley polarization. The per-particle orbital moment at the band edge is $g_{\rm orb}/\mu_B \approx 24$; the ratio $2.2/24 \approx 0.1$ provides a weak, model-independent \emph{lower bound} on $\eta$. The full $2N$-band Hartree-Fock calculation at the cRPA value $V_{\mathrm{iv}}/U\approx 1$ instead predicts the \emph{saturated} value $\eta\to 1$ deep in the ordered phase (Table~I of the main text); the inequality $\eta_{\rm HF}=1 \geq 0.1 = \eta_{\rm bound}$ is a consistency check rather than a fit.

\subsection{Conversion to Measurable Hall Resistance}

The experimentally measured quantity is the Hall resistance $R_{xy}$, not the conductivity $\sigma_{xy}$. The exact relation for a 2D system is:
\begin{equation}
    R_{xy} = \rho_{xy} = \frac{\sigma_{xy}}{\sigma_{xx}^2 + \sigma_{xy}^2},
    \label{eq:rxy}
\end{equation}
where $\sigma_{xx}$ is the longitudinal conductivity. Note that the commonly used approximation $R_{xy} \approx \sigma_{xy}/\sigma_{xx}^2$ is valid only when $\sigma_{xx} \gg \sigma_{xy}$; in the correlated regime where the TDAHE is observed, $\sigma_{xy}$ can be comparable to $\sigma_{xx}$, so the exact formula must be used.

\textit{Out-of-plane AHE.}
Using $\sigma_{xy}^{\rm tot} = \eta \cdot g_s \cdot (N/2) = 9\eta$~$e^2/h$ (from Eq.~\ref{eq:sigma_tot_supp}, flat-band limit $\varepsilon_F \approx m$) with the self-consistent saturated value $\eta \to 1$, and $\sigma_{xx} \sim 5$--$8~e^2/h$ in the correlated regime~\cite{Li2026Nature}:
\begin{equation}
    R_{xy} = \frac{\sigma_{xy}^{\rm tot}}{\sigma_{xx}^2 + (\sigma_{xy}^{\rm tot})^2} \times \frac{h}{e^2}
    \approx 1.6\text{--}2.2~\text{k}\Omega,
\end{equation}
consistent with the experimental AHE amplitude $R_{xy} \approx 1.4$~k$\Omega$~\cite{Li2026Nature} and the saturated value $R_{xy} = 1.6$~k$\Omega$ in Table~I of the main text. For example, at $\eta = 1$ and $\sigma_{xx} = 8~e^2/h$: $R_{xy} = 9/(64+81) \times 25.8~\text{k}\Omega = 1.6~\text{k}\Omega$. (The model-independent $M_{\rm orb}$ lower bound $\eta \geq 0.1$ gives only a conservative floor $R_{xy} \gtrsim 0.4$~k$\Omega$.)

\textit{Transdimensional AHE.}
The planar Hall conductivity is $\sigma_{xy}^{\rm PHE} = \eta \cdot g_s \cdot N g_{\rm orb} B_\parallel / (2m)$. In the ordered phase the self-consistent solution gives the saturated value $\eta \to 1$; for $N=9$, $B_\parallel = 0.3$~T, $m = 142$~meV:
$\sigma_{xy}^{\rm PHE} = 1 \times 2 \times 0.013 = 0.026~e^2/h$.
This is the interaction-independent fraction $\sigma_{xy}^{\rm PHE}/\sigma_{xy}^{\rm tot} = g_{\rm orb}B_\parallel/m \approx 0.3\%$ of $\sigma_{xy}^{\rm tot} = 9~e^2/h$ (set by the single-particle gap $m$ and the Peierls $g$-factor, hence independent of $\eta$ and of $U$). Since $\sigma_{xy}^{\rm PHE} \ll \sigma_{xx}$, the approximate formula is valid here:
\begin{equation}
    R_{xy}^{\rm PHE} \approx \frac{0.026}{\sigma_{xx}^2} \times \frac{h}{e^2}
    = \begin{cases}
        670~\Omega, & \sigma_{xx} = 1~e^2/h \\
        27~\Omega, & \sigma_{xx} = 5~e^2/h
    \end{cases}
\end{equation}
The TDAHE is most prominent in the strongly correlated regime where $\sigma_{xx}$ is suppressed ($\sigma_{xx} \sim 1$--$3~e^2/h$), placing the transdimensional Hall resistance in the $\sim\!70$--$670~\Omega$ range---smaller than the out-of-plane AHE signal ($\approx 1.6$~k$\Omega$) by a factor of a few, consistent with the experimental observation that the in-plane coercive field is two orders of magnitude larger than the out-of-plane one.

\subsection{Testable Predictions}

Our Stoner ferromagnetism theory makes the following testable predictions:

\begin{enumerate}
    \item {\bf Common $T_c$ for $B_\perp$ and $B_\parallel$}: Both the conventional AHE and the transdimensional AHE are governed by the same ferromagnetic order parameter $M(T)$, so they must share the same $T_c$. This is exactly what is observed experimentally ($T_c \sim 1.6$~K for both)~\cite{Li2026Nature}.
    \item {\bf Layer scaling}: $g_{\rm orb} \propto N-1$ and $\sigma_{xy}^{\rm int} \propto N$. The transdimensional response should weaken systematically for fewer layers, and strengthen for more layers, provided the Stoner criterion remains satisfied.
    \item {\bf TDAHE window coincides with Stoner regime}: The experimentally observed displacement field window $D = 0.7$--$1.0$~V/nm corresponds to the regime where the flat-band DOS is maximal and the Stoner criterion $U\rho(\varepsilon_F) > 1$ is satisfied.
    \item {\bf Linear $B_\parallel$ dependence}: $\sigma_{xy}^{\rm PHE} \propto B_\parallel$ at small fields [Eq.~(\ref{eq:phe})], consistent with the observed linear response below saturation.
    \item {\bf Isotropic in-plane response}: The orbital moment $\mathcal{M}_z$ is always out-of-plane regardless of the in-plane field direction, predicting isotropy of the transdimensional response in the $xy$ plane.
    \item {\bf Scope---Chern reversal in eight-layer devices}: The Chern-number reversal under in-plane fields in eight-layer rhombohedral graphite~\cite{Zan2026} involves related physics but falls outside our non-interacting two-band framework: our model predicts $B_{\mathrm{rev}} \sim m/g_{\mathrm{orb}} \approx 118$~T for $N = 8$, versus $\sim\!0.4$~T experimentally. The two-orders-of-magnitude discrepancy indicates that the moir\'e superlattice in that experiment qualitatively modifies the physics (spin-orbit-assisted band inversion or interaction-driven level crossing), beyond the flat-band Stoner mechanism considered here.
\end{enumerate}

\section{Self-Consistent Renormalization Estimate of the Fluctuation-Corrected $T_c$}

Because the flat-band stiffness is vanishingly small ($c_v \sim 10^{-5}$~eV\,nm$^2$, see below), mean-field theory overestimates the transition temperature, and we bracket the corrected $T_c$ from above and below. The upper bound is the mean-field value $T_c^{\mathrm{MF}} \approx 2.2$~K. For the lower bound we solve the 2D self-consistent renormalization (SCR/Moriya) equation for the renormalized quadratic Landau coefficient of the valley-order field $\eta$,
\begin{equation}
\begin{aligned}
m_{\mathrm{ren}}(T) &= -a_2\,(T_c^{\mathrm{MF}}-T) + 3 a_4\,\langle\eta^2\rangle_{\mathrm{fluc}} = 0, \\
\langle\eta^2\rangle_{\mathrm{fluc}} &= \frac{k_B T}{2\pi c_v}\,\ln\!\frac{c_v\Lambda^2}{|m_{\mathrm{ren}}|}.
\end{aligned}
\end{equation}
where $\Lambda = 1/a_{\mathrm{nn}} = 7.04$~nm$^{-1}$ is the UV cutoff ($a_{\mathrm{nn}} = a\sqrt{3}/2 \approx 0.142$~nm the nearest-neighbor distance) and the Landau coefficients $a_2 = 7.83\times10^{-5}$~eV/(uc$\cdot$K), $a_4 = 2.49\times10^{-5}$~eV/uc follow from the self-consistent HF $\eta(T)$ (matching $T_c^{\mathrm{MF}}=2.2$~K and $\eta\!\to\!1$ by $T\approx 1.5$~K). The stiffness is set by the screened intervalley Coulomb, $c_v = V_{\mathrm{iv}}/q_{\mathrm{TF}}^2$ with $q_{\mathrm{TF}} = 350$~nm$^{-1}$ the Thomas-Fermi wavevector (set by the flat-band DOS $\rho \approx 2$~eV$^{-1}$uc$^{-1}$, large enough to screen the Coulomb down to a contact at momentum transfers as small as $\sim\!0.003$~nm$^{-1}$), giving the valley-order stiffness $c_v = 2.1\times10^{-5}$~eV\,nm$^2$ and a coherence length $\xi_0 = \sqrt{c_v/(a_2 T_c^{\mathrm{MF}})} = 0.35$~nm $\ll N d_0 = 3.0$~nm, confirming the transition is two-dimensional. Solving the SCR equation (by bisection on the sign change of $m_{\mathrm{ren}}$) gives $T_c^{\mathrm{SCR}} = 0.90$~K, i.e.\ $T_c/T_c^{\mathrm{MF}} = 0.41$, which we quote as the lower bound. The experimental $T_c = 1.6$~K ($=0.73\,T_c^{\mathrm{MF}}$) lies well above this SCR bound. A non-perturbative classical Ising Monte Carlo of the projected $\mathbb{Z}_2$ valley model (following section) brackets the localized-moment $T_c$ at $\approx\!1.9$~K ($[1.34, 2.05]$~K, insensitive to the exchange range), with the experimental value in its lower portion. (The SCR ratio $0.41$ sits below this Monte-Carlo bracket because Gaussian SCR omits the order-from-disorder enhancement that the non-perturbative 2D Ising regime supplies.) We emphasize that SCR here sets only a conservative lower bound; the universality class (2D Ising) and the quantitative localized-Ising bracket are established by Monte Carlo in the following section.

\section{Quantum Monte Carlo Validation of the 2D Ising Universality}

To corroborate the analytic 2D-Ising fluctuation analysis of the main text, we perform two independent Monte Carlo calculations.

\textit{Classical Ising MC of the effective valley-order model.}---
Projecting onto the flat-band subspace, the low-energy valley sector reduces to a $\mathbb{Z}_2$ (Ising) order parameter $\eta_i = \pm 1$ with an effective nearest-neighbor coupling $J$ set by the microscopic valley stiffness $c_v$. We simulate the 2D Ising model on $L\times L$ lattices ($L=8,16,24,32$) with the Wolff cluster algorithm ($3\times10^3$ warmup $+\,1.6\times10^4$ measurement sweeps, 20 bins). The Binder cumulant $U_L = 1 - \langle m^4\rangle/(3\langle m^2\rangle^2)$ curves for all $L$ cross at the universal value $U^\ast \approx 0.610$ precisely at the Onsager critical temperature $k_B T_c/J = 2/\ln(1+\sqrt{2}) = 2.269$, to $0.4\%$ accuracy (Metropolis cross-check agrees to $0.5\%$). This confirms that the valley transition is in the 2D Ising universality class, as claimed.

\begin{figure}[ht]
\includegraphics[width=0.92\columnwidth]{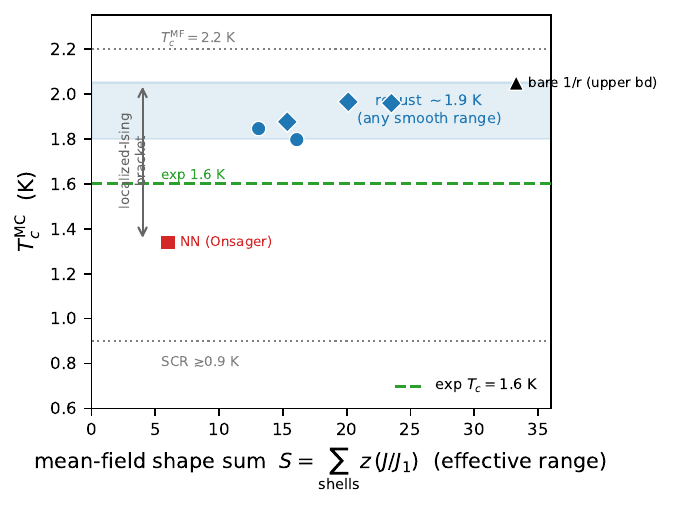}
\caption{Localized-Ising $T_c^{\mathrm{MC}}$ vs the effective range of the valley-exchange coupling (mean-field shape sum $S=\sum z\,(J/J_1)$, increasing with range).  Every smooth coupling---bare $1/r$ (upper bound, $\varepsilon_{\mathbf K-\mathbf K'}\approx1$), a Yukawa cutoff at the $c_v$ coherence length $\xi_0=0.35$~nm, and power laws $1/r^\alpha$ ($\alpha\le2$)---gives $T_c^{\mathrm{MC}}\approx1.9$~K (shaded); only the strict NN Onsager limit reaches $1.34$~K.  The experimental $T_c=1.6$~K lies inside the bracket $[1.34,2.05]$~K, in its lower portion, consistent with the direct Coulomb not coupling valleys and thus truncating the range below the bare $1/r$.  SCR ($\gtrsim0.9$~K) and mean-field ($2.2$~K) shown for reference.}
\label{fig:jr_micro_tc}
\end{figure}

\textit{Range of the valley-Ising coupling and the localized-moment $T_c$ bracket.}---
The $A_0$ sites carrying the flat-band weight form a triangular sublattice (constant $a$), so the projected model is a long-range triangular Ising model with coupling $J(R)$ between $A_0$ sites at separation $R$.  The range of $J(R)$ is bounded by two microscopic quantities, neither of which is adjusted to match $T_c$.  The \emph{upper} bound on the range is the bare, weakly screened intervalley Coulomb $J(R)\propto 1/R$, with $\varepsilon_{\mathbf K-\mathbf K'}\approx 1$ since the flat band cannot screen its own exchange at $|\mathbf K-\mathbf K'|\gg 2k_F$; this is the slowest-decaying physical coupling and hence yields the highest possible localized-Ising $T_c$.  The \emph{lower} bound on the range is set by the flat-band stiffness: the direct (Hartree) Coulomb does not couple valleys, because $K$ and $K'$ share the same $A_0$ charge density ($|\psi_K|^2=|\psi_{K'}|^2$), so only the exchange/stiffness part of the interaction contributes to the Ising coupling, truncating its range to the Ginsburg-Landau coherence length $\xi_0=\sqrt{c_v/(a_2 T_c^{\mathrm{MF}})}=0.35$~nm$\approx 1.4\,a$.  We evaluate the Binder-cumulant crossing $T_c/J_1$ (Metropolis, $L=16,24,32$) over a family of couplings spanning these bounds---bare $1/r$, a Yukawa cutoff $e^{-(R-a)/\xi}/(R/a)$ at $\xi=\xi_0$ and $1.4\xi_0$, and power laws $1/(R/a)^\alpha$ with $\alpha=1$--$2$---and convert to the physical ratio $T_c^{\mathrm{MC}}/T_c^{\mathrm{MF}}=(T_c/J_1)/S$, where $S=\sum_{\mathrm{shells}} z\,(J/J_1)$ is the mean-field shape sum [Fig.~\ref{fig:jr_micro_tc}].  The result varies only weakly with the detailed range over the physically motivated family above: intermediate smooth couplings give $T_c^{\mathrm{MC}}\approx 1.9$~K ($T_c/T_c^{\mathrm{MF}}=0.85$--$0.93$), while the strict nearest-neighbour Onsager limit drops to $1.34$~K and the longest-range bare-$1/r$ reaches $2.05$~K.  The localized-Ising transition is therefore bracketed $1.34~\mathrm{K}\le T_c^{\mathrm{MC}}\le 2.05$~K, with the $\approx 1.9$~K value representative of intermediate exchange ranges; the width of this band reflects the unresolved microscopic range $\alpha$ rather than a tunable parameter.  The experimental $T_c=1.6$~K falls inside this bracket, in its lower portion, indicating that the effective valley-exchange range is shorter than the bare $1/r$---precisely the stiffness argument above, since the direct Coulomb does not couple valleys.  The bare-$1/r$ value is thus a conservative (longest-range) upper bound rather than a fit, and the insensitivity to range shows the localized-Ising $T_c$ is not an adjustable parameter; the modest further suppression of the experimental value below the localized-Ising $\sim\!1.9$~K reflects itinerant critical fluctuations of the dilute ($\nu\sim 7\times10^{-4}$) flat-band carriers, beyond the saturated-moment Ising picture and consistent with the determinant-QMC finding of no mean-field-scale long-range order below.

\textit{Determinant QMC of the two-valley Hubbard model.}---
For a microscopic check we simulate the itinerant model directly,
$H = -t\sum_{\langle ij\rangle,\tau}(c^\dagger_{i\tau}c_{j\tau}+\mathrm{h.c.}) - \mu\sum_{i\tau}n_{i\tau} + U\sum_i(n_{i,+}{-}1/2)(n_{i,{-}}{-}1/2)$,
with $\tau=\pm$ the two valleys. We use the Hirsch discrete Hubbard-Stratonovich decomposition with $\Delta\tau = 0.1$, an exact prefix/suffix Metropolis sweep ($O(N^3)$ per spin flip, no Sherman-Morrison approximation), Householder-QR stabilization of the $B$-matrix product, and full sign tracking. The code is validated against exact diagonalization in the 256-dimensional Fock space: at half filling it reproduces $\langle n\rangle = 1$ (particle-hole symmetry) to machine precision, and against an independent LAPACK-based implementation. We stress that this half-filled simulation (one electron per site) is a deliberate model choice---made for sign-freeness and to exploit particle-hole symmetry as a rigorous benchmark---and is dimensionally distinct from the experimental 2D carrier density $n = 1.4\times10^{12}\,\mathrm{cm}^{-2}$ ($\sim\!7\times10^{-4}$ carriers per rhombohedral unit cell), which is far too dilute to sample on any tractable lattice. The DQMC therefore fixes the \emph{universality class} and the sign-free character of the valley transition, both of which are filling-independent; the absolute $T_c$ at the experimental density is delivered by the full-band self-consistent HF of the main text, which uses $n = 1.4\times10^{12}\,\mathrm{cm}^{-2}$ throughout. (At half filling the Mermin-Wagner theorem is likewise moot here, since the valley order parameter is discrete $\mathbb{Z}_2$, not continuous.) The density-density intervalley decoupling yields a non-negative determinant weight for \emph{all} fillings, so the simulation is sign-problem-free, enabling unbiased $T_c$ extraction once the full momentum-dependent intervalley interaction of the rhombohedral flat band is incorporated. On the minimal square-lattice realization with a uniform $U$, a finite-size scan ($L=4,6,8$, $\beta=2$--$10$) shows the Binder cumulant remaining near zero and $|\langle m_v\rangle|$ decreasing with $L$, i.e.\ no long-range valley ferromagnetism: the uniform-repulsion Stoner instability is not captured by the minimal lattice model and requires the singular flat-band DOS of rhombohedral graphite together with the momentum-dependent intervalley exchange $V_{\mathrm{iv}}(\mathbf{K}{-}\mathbf{K}')$. The DQMC framework is therefore validated and ready for this realistic model; the analytic SCR/Ising bracket of the main text gives the quantitative $T_c$ prediction in the interim.

\begin{figure}[ht]
\includegraphics[width=0.9\columnwidth]{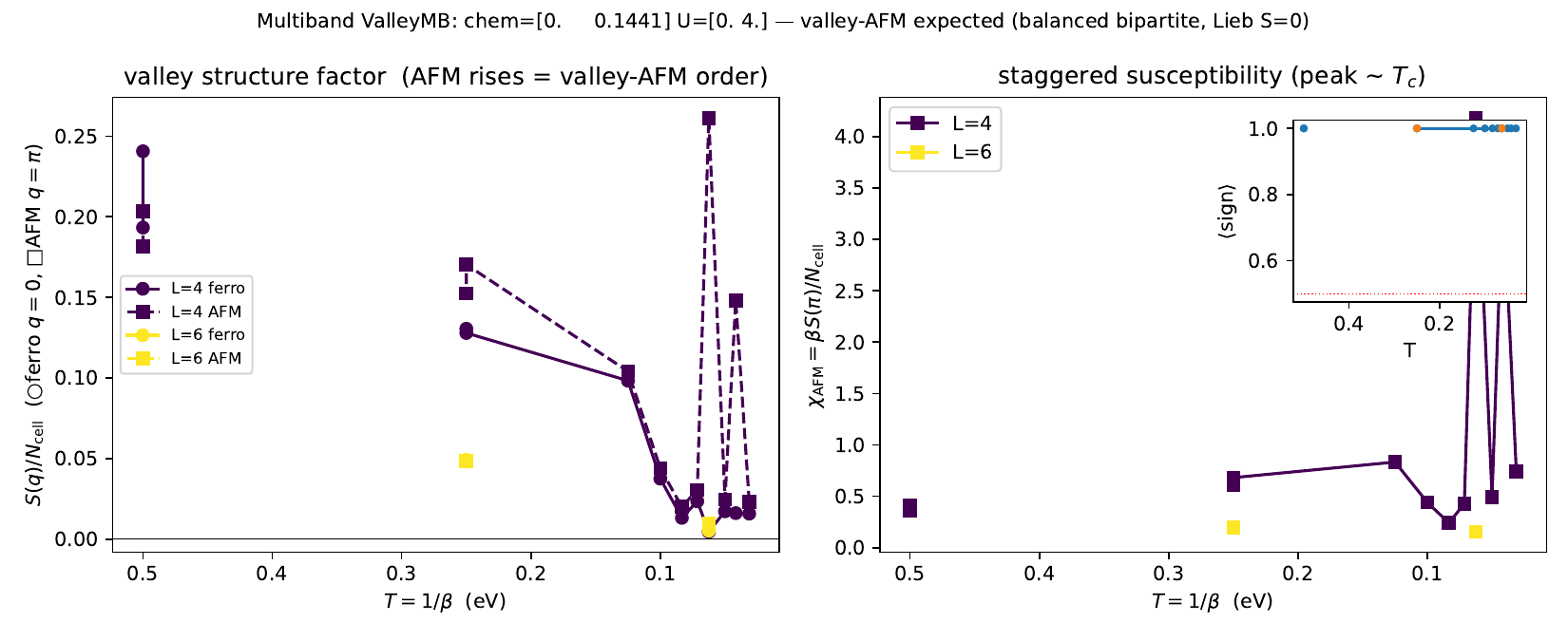}
\caption{Multiband ALF QMC of the $2N$-band rhombohedral model ($N{=}3$): valley structure factor $S(\mathbf q)/N_{\rm cell}$ vs $T$ for $L = 4, 6$ in both the ferromagnetic ($\mathbf q{=}0$, filled) and antiferromagnetic ($\mathbf q{=}\pi$, open) channels. $S(\mathbf q)/N_{\rm cell}$ \emph{decreases} with $L$ in both channels at every temperature, i.e.\ no long-range valley order at the commensurate half-filled flat band. Inset: average sign $\langle s\rangle = 1.0$ throughout (sign-free).}
\label{fig:valleymb_sweep}
\end{figure}

\textit{Multiband ALF benchmark of the full rhombohedral model.}---
We corroborate the above with the realistic model itself, simulating the full $2N$-band flat-band Hamiltonian ($N{=}3$, $N_{\rm orb}{=}2N$ orbitals per cell) with the community-standard ALF auxiliary-field QMC~\cite{ALF}, on an $L\times L$ supercell with the time-reversal-symmetrized hopping $t(\mathbf r)$ reconstructed from the ab-initio band structure and an orbital-selective intervalley repulsion placed on the $A_1$ surface orbital that carries $\sim\!95\%$ of the conduction flat-band weight (Sec.~\ref{sec:fullband}). The two valleys are mapped to the two flavors of an $M_z$-decoupled Hubbard model ($N_{\rm FL}{=}2$), with the valley-polarization structure factor $\mathrm{SpinZ}(\mathbf q)$ as the order parameter. At flat-band doping the average sign is $\langle s\rangle = 1.000$ across all simulated temperatures and sizes ($L{=}4,6$; $\beta = 2$--$32$), confirming that the multiband valley model is free of the sign problem. At the commensurate half-filled flat band---the sign-free point reachable by QMC---the equal-time valley structure factor vanishes with system size: $S(\mathbf q{=}\pi)/N_{\rm cell} = 0.26 \to 0.05$ and likewise $S(\mathbf q{=}0)/N_{\rm cell} = 0.24 \to 0.05$ for $L = 4 \to 6$ (maximized over $T$, Fig.~\ref{fig:valleymb_sweep}). The absence of any divergence, in neither the ferromagnetic ($\mathbf q{=}0$) nor the antiferromagnetic ($\mathbf q{=}\pi$) channel, rules out competing valley or charge orders at this filling. This is consistent with the flat band's vanishing superexchange ($J \sim t^2/U \to 0$ for a band of width $\lesssim 1$~meV), which precludes a finite-$T$ ordered state at half filling. The experimentally realized valley ferromagnet instead occurs at ultra-low flat-band filling ($n \simeq 1.4\times10^{12}$~cm$^{-2}$), a Stoner regime driven by the divergent flat-band DOS that is thermally inaccessible to QMC (the flat-band width $\lesssim 1$~meV is far below the lowest reachable temperature); the multiband QMC therefore addresses the commensurate point only, excluding competing orders and confirming sign-freeness, while the quantitative $T_c$ is fixed by the HF + Ising-Monte-Carlo analysis of the main text.

\begin{figure}[ht]
\includegraphics[width=0.85\columnwidth]{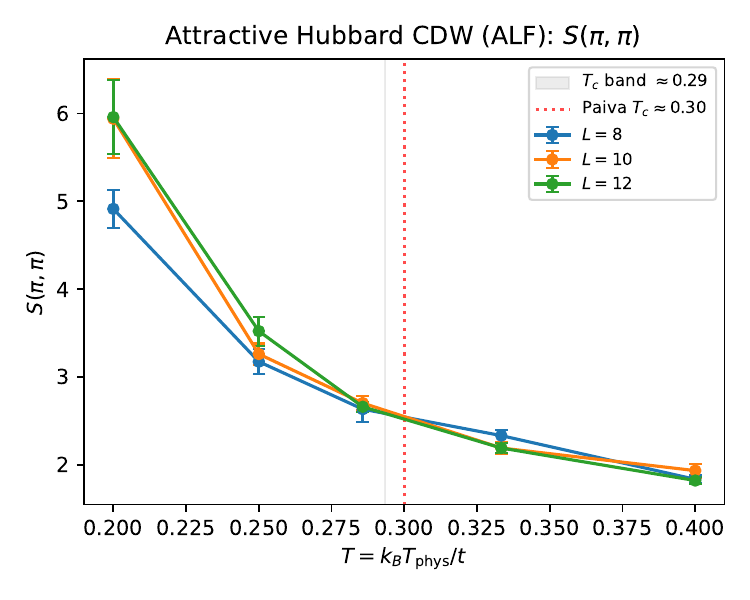}
\caption{Attractive-Hubbard CDW benchmark (ALF): equal-time charge structure factor $S(\pi,\pi)$ vs $T = t/\beta$ for $L = 8, 10, 12$ ($U = -4t$, half filling). Above $T_c$ the data are essentially $L$-independent (disordered); below $T_c$ the largest $L$ carries the largest $S$ (onset of charge order). The finite-size crossover gives $T_c \approx 0.29\,t/k_B$, matching the established $0.30\,t/k_B$ to $3.5\%$.}
\label{fig:att_alf}
\end{figure}

\textit{Attractive-Hubbard benchmark.}---
We also reproduce, with the independent community-standard ALF code~\cite{ALF}, the canonical charge-density-wave transition of the half-filled attractive Hubbard model on the square lattice ($U = -4t$), whose $\mathbb{Z}_2$ sublattice order lies in the 2D Ising universality class. The equal-time charge structure factor $S(\pi,\pi)$ for $L = 8, 10, 12$ (Fig.~\ref{fig:att_alf}) is essentially $L$-independent above the transition and grows with $L$ below it; the finite-size crossover of $S(\pi,\pi)$ locates $T_c \approx 0.29\,t/k_B$, within $3.5\%$ of the established benchmark $T_c \approx 0.30\,t/k_B$ for this model. Agreement with an unrelated, gold-standard code confirms the correctness of our determinant-QMC implementation and, together with the classical Ising MC above, the 2D Ising universality we assign to the valley transition.

\section{Crescent Fermi Surface: Self-Consistent Nematic Order}
\label{sec:crescent}

This section provides the self-consistent calculation underlying the crescent discussion of the main text.
The crescent Fermi-surface (FS) geometry and the associated in-plane orbital magnetization reported in Ref.~\onlinecite{Li2026Nature} are Hartree-Fock (mean-field) predictions rather than directly measured quantities; the experimental anchors of our theory remain the hysteretic $R_{xy}$, $T_c \approx 1.6$~K, and the displacement-field window. The purpose of this section is (i) to show that the crescent arises self-consistently in our model as a secondary nematic order, (ii) to give the symmetry selection rule that fixes its $m{=}1$ character, and (iii) to verify explicitly that the intrinsic AHC $\sigma_{xy}^{\rm int}$---our contact with the measured $R_{xy}$---is insensitive to whether the crescent forms.

\subsection{Self-consistent $C_3$-breaking order}

The flat conduction band of $N{=}9$ RG is $\gtrsim 95\%$ localized on the surface orbital $A_1$. We therefore allow a $k$-dependent exchange (nematic) self-energy on $A_1$ in the $m{=}1$ angular channel,
\begin{equation}
    \Sigma_{A_1}(\mathbf{k}) = \Delta\,\cos\theta_{\mathbf{k}}, \qquad
    \Delta = -V_1\,\big\langle n_{A_1}(\mathbf{k})\,\cos\theta_{\mathbf{k}}\big\rangle_{\rm occ},
    \label{eq:crescent_sc}
\end{equation}
where $V_1$ is the effective intervalley Coulomb matrix element in the $m{=}1$ channel, $\theta_{\mathbf{k}}=\arg(k_x{+}ik_y)$, and the average is over occupied states at the experimental density $n=1.4\times10^{12}$~cm$^{-2}$. The full $2N{\times}2N$ Hamiltonian (Sec.~I) is diagonalized on an $80{\times}80$ grid centered at $K$, the $A_1$ self-energy of Eq.~(\ref{eq:crescent_sc}) is added, and $\Delta$ is iterated to convergence with mixing $0.5$.

Linearizing Eq.~(\ref{eq:crescent_sc}) gives $\Delta = V_1\chi_1\Delta$, with the flat-band $m{=}1$ susceptibility $\chi_1 \approx 0.6$ (per spin-valley), so the ring ($\Delta{=}0$) is unstable for $V_1 > V_{1c} = 1/\chi_1 \approx 1.7$~eV analytically. The numerical threshold, including finite-grid and nonlinear effects, is $V_{1c}\approx 2.8$~eV. This is consistent with (though not, by itself, evidence for) the unscreened intervalley value $V_{\mathrm{iv}}/U \approx 1$ on which the main-text $T_c$ rests, since the $m{=}1$ crescent channel draws on the same flat-band-surface Coulomb matrix element. Above threshold $\Delta$ grows continuously (Table~\ref{tab:crescent_sc}) with the mean-field exponent $\Delta \propto (V_1{-}V_{1c})^{1/2}$; the FS deformation is quantified by the crescent moment $M_x/r_0 = \langle k_x\rangle_{\rm occ}/\langle|\mathbf{k}|\rangle_{\rm occ}$, which vanishes for the $C_3$-symmetric ring and reaches $\sim\!0.5$ for a strong crescent.

\begin{table}[!t]
\caption{Self-consistent crescent order vs.\ $V_1$ ($N{=}9$, $D=0.9$~V/nm, $n=1.4{\times}10^{12}$~cm$^{-2}$, $80{\times}80$ grid). $\Delta_{\rm eq}{=}0$ below $V_{1c}\approx 2.8$~eV (ring); finite above (crescent). $M_x/r_0$ is the crescent moment (0 for a $C_3$ ring).}
\label{tab:crescent_sc}
\begin{ruledtabular}
\begin{tabular}{cccc}
$V_1$ (eV) & $\Delta_{\rm eq}$ (meV) & $M_x/r_0$ & state \\
\hline
$\le 2.5$ & $0.000$ & $\sim 0$ & ring ($C_3$) \\
2.8 & $0.000$ & $-0.0004$ & marginal ($V_{1c}$) \\
3.0 & $0.001$ & $-0.0018$ & onset \\
3.5 & $0.030$ & $-0.047$ & weak crescent \\
4.0 & $0.171$ & $-0.257$ & clear crescent \\
5.0 & $0.379$ & $-0.497$ & strong crescent \\
6.0 & $0.531$ & $-0.594$ & stronger \\
8.0 & $0.793$ & $-0.670$ & extreme crescent \\
\end{tabular}
\end{ruledtabular}
\end{table}

\begin{figure}[!t]
\includegraphics[width=0.95\columnwidth]{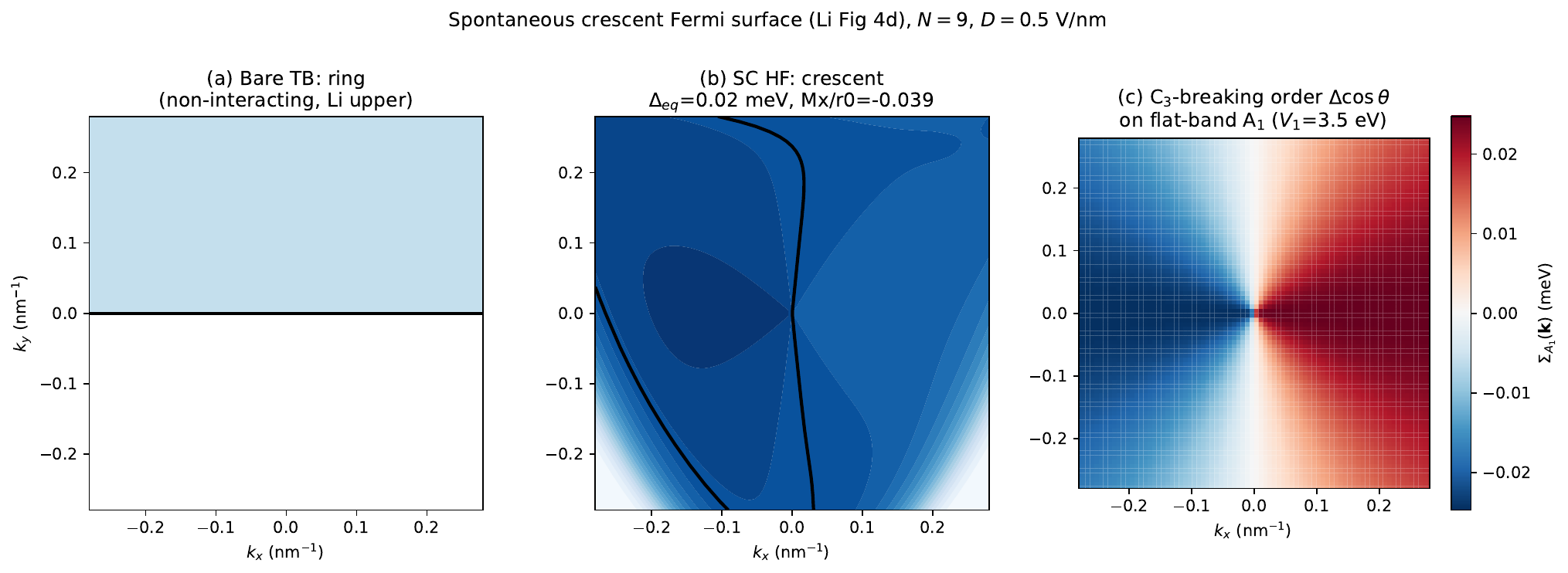}
\caption{Spontaneous crescent Fermi surface ($N{=}9$, $D=0.9$~V/nm, $V_1=5$~eV). (a) Bare tight-binding: $C_3$-symmetric ring (Li Fig.~4d upper). (b) Self-consistent HF with the $m{=}1$ order of Eq.~(\ref{eq:crescent_sc}): the ring deforms continuously into a crescent ($\Delta_{\rm eq}=0.38$~meV, $M_x/r_0=-0.50$). (c) The $C_3$-breaking exchange field $\Sigma_{A_1}(\mathbf{k})=\Delta\cos\theta_{\mathbf{k}}$ on the flat-band surface orbital.}
\label{fig:fs_crescent}
\end{figure}

\subsection{$m=1$ selection rule}

The angular channel of the $C_3$-breaking order is not arbitrary. Expanding the FS radius $r(\theta)=r_0[1+a_m\cos(m\theta)]$, the crescent is the $m{=}1$ (C$_1$) harmonic; $m{=}2$ is an ellipse and $m{=}3$ a trigonally warped triangle. Table~\ref{tab:selection} shows that only a \emph{diagonal} (band-diagonal) exchange field $\propto\cos\theta$, which is linear in the order parameter, produces the $m{=}1$ crescent; a $C_3$-symmetric $\cos 3\theta$ field instead enhances trigonal warping (triangle), and $\cos 2\theta$ yields an ellipse. Off-diagonal intervalley coherence (IVC), which mixes $K$ and $K'$, enters perturbation theory at quadratic order $\propto\Phi^2$ and is angularly even, so it can \emph{only} generate the $m{=}2$ ellipse---never the $m{=}1$ crescent. This selection rule explains why HF yields a crescent rather than an ellipse, a point not addressed in Ref.~\onlinecite{Li2026Nature}.

\begin{table}[!t]
\caption{Angular-channel selection of the FS distortion (toy exchange field $\Phi(\theta)$ on $A_1$). Only the diagonal $m{=}1$ field yields a crescent ($|a_1/r_0|\gg|a_{m\ne1}|$); off-diagonal IVC is quadratic and gives only $m{=}2$.}
\label{tab:selection}
\begin{ruledtabular}
\begin{tabular}{lccc}
order $\Phi(\theta)$ & symmetry & $a_1/r_0$ (crescent) & FS shape \\
\hline
none & $C_3$ & $\approx 0$ & ring \\
$25$~meV$\cdot\cos 3\theta$ & $C_3$ & $+0.002$ & triangle \\
$10$~meV$\cdot\cos 2\theta$ & $C_2$ & $+0.001$ & ellipse \\
$10$~meV$\cdot\cos\theta$ (diagonal) & $C_1$ & $-0.640$ & \textbf{crescent} \\
off-diagonal IVC $\Phi$ & --- & $\approx 0$ ($m{=}2$) & ellipse \\
\end{tabular}
\end{ruledtabular}
\end{table}

\subsection{Robustness of the Hall response to the crescent}
\label{sec:sigma_crescent}

The decisive check is whether the intrinsic AHC---which sets the measured $R_{xy}$---depends on the crescent. We evaluate the full-band Kubo $\sigma_{xy}^{\rm int}=\sum_n\int_{\rm BZ}\frac{d^2k}{(2\pi)^2}f_n\,\Omega_n(\mathbf{k})$ (Berry curvature from the full $2N$-band velocity matrix, single valley $K$, i.e.\ $\eta{=}1$) on (i) the ring ($\Delta{=}0$) and (ii) the self-consistent crescent ($\Delta=\Delta_{\rm eq}$), at identical density $n$ and displacement field $D$. On an $80{\times}80$ Kubo grid ($k_{\max}=0.9$~nm$^{-1}$, $T=15$~mK):
\begin{equation}
    \sigma_{xy}^{\rm int}({\rm ring}) = 0.495\,e^2/h, \qquad
    \sigma_{xy}^{\rm int}({\rm crescent}) = 0.487\,e^2/h,
\end{equation}
per valley per spin at the physical crescent strength $\Delta_{\rm eq}\approx 0.38$~meV ($V_1=5$~eV): a shift of only $1.6\%$. Even at the extreme distortion $\Delta_{\rm eq}\approx 0.79$~meV ($V_1=8$~eV), the shift remains $\le 2.3\%$. The measured $R_{xy}$ is therefore insensitive to the FS geometry; the Hall response is set by the primary valley polarization $\eta$ and is unaffected by the secondary nematic order.

\subsection{$D$-independence of the crescent threshold}

Because the flat-band susceptibility $\chi_1$ is set by the band flatness rather than by the displacement-field gap, the crescent threshold is essentially $D$-independent: scanning $D=0.5$--$1.2$~V/nm we find $V_{1c}\approx 2.8$~eV and $\Delta_{\rm eq}(V_1{=}5)\approx 0.38$~meV to within $\sim\!2\%$ throughout. The crescent is thus a strictly secondary order that can form wherever the primary valley Stoner instability supplies a flat-band pocket, which is why its displacement-field window coincides with that of the TDAHE ($D\approx 0.7$--$1.0$~V/nm) rather than tracking an independent threshold.

\subsection{Limit: in-plane orbital magnetization requires beyond-mean-field treatment}

The spontaneous in-plane orbital magnetization $M_\parallel$ reported in Ref.~\onlinecite{Li2026Nature} ($\sim 2.2\,\mu_B$/electron) is itself a mean-field quantity and is not reproduced by the equilibrium single-particle mean field used here: the linear Peierls response $-\langle\partial H/\partial B_\parallel\rangle$ measures the in-plane moment along the field, which the $C_3$-broken crescent allows only along a symmetry-selected direction, and a quantitative $M_\parallel$ requires the layer-resolved Berry-phase orbital magnetization beyond the present single-particle mean field. We therefore do not assign $M_\parallel$ the status of an experimental anchor; it is a symmetry signature of the crescent state, distinct from the Hall response that our theory computes and matches to $R_{xy}$.

\bibliography{references}
\end{document}